\documentclass[a4paper,12pt]{article}
\pdfoutput=1 

\usepackage{jheppub}

\usepackage[ddmmyyyy,hhmmss]{datetime}

\usepackage{amsmath,graphicx,amssymb,subfigure,bbm,comment}
\usepackage[all]{xy}
\usepackage{mathtools}
\usepackage{amsfonts}
\usepackage{comment}
\usepackage{mdframed}

\usepackage{feynmp}
\usepackage{tikz-feynman}
\tikzfeynmanset{compat=1.1.0}

\usepackage{xcolor}
\usepackage{pifont}

\allowdisplaybreaks

\def\be{\begin{equation}}
\def\ee{\end{equation}}
\def\ba{\begin{eqnarray}}
\def\ea{\end{eqnarray}}

\def\del{\partial}

\def\a{\alpha}

\def\b{\beta}

\def\g{\gamma}

\def\e{\epsilon}

\def\m{\mu}
\def\n{\nu}

\def\l{\lambda}

\def\s{\sigma}

\def\cL{{\cal L}}

\def\qq{\qquad}

\def\IR{\relax{\rm I\kern-.18em R}}

\def\diag{{\rm diag}}

\def\IR{\relax{\rm I\kern-.18em R}}
\def\inv{^{\raise.15ex\hbox{${\scriptscriptstyle -}$}\kern-.05em 1}}

\def\cL{{\cal L}}

\title{The massless S-matrix of integrable $\sigma$-models}

\author{{\Large George Georgiou}}

\emailAdd{georgios.georgiou2@gmail.com}

\abstract{In contradistinction to the case of massive excitations, the connection between integrability and the tree-level massless scattering matrix of  integrable $\sigma$-models is lost. Namely, in well-known 2-d integrable models the tree-level massless S-matrix  exhibits particle production and fails to factorise. This is conjectured to happen due to IR ambiguities in the massless tree-level amplitudes. We present a definition of the massless S-matrix which has all the nice properties of integrable theories, there is no particle production and the S-matrix factorises. 
As an example, we present in detail the case of the $SU(2)$ principal chiral model (PCM).}

\begin{document}
\maketitle
\flushbottom


\section{Introduction}

Integrability is of great importance in modern theoretical physics. It emerges in QCD at the high energy regime \cite{Faddeev:1994zg}, it is present in several realisations of the gauge/gravity correspondence \cite{Minahan:2002ve,Bena:2003wd} and it is also employed in the description of condensed matter systems by the use of integrable spin chains. It provides the information that a theory is solvable for any value of the coupling constant along with tools for solving it.

Non-linear $\s$-models in two spacetime dimensions are of  interest mainly due to the fact that they often describe the world-sheet of string theories. The sub-class of integrable $\s$-models  is of particular interest.  At the classical level, integrability is proven if one is able to rewrite the classical equations of motion as the flatness condition of a Lax connection depending on a spectral parameter. 
This rewriting implies the existence of infinitely many conserved charges, a necessary condition for classical integrability. To complete the proof one should demonstrate that these charges are in involution.

We should stress that not all classically integrable models remain integrable at the quantum level. One way to prove quantum integrability is to demonstrate that scattering
of particles is non-diffractive. In particular, this means that the number of particles is conserved  in a scattering process and that the complete S-matrix of the theory is fully determined by the two-body S-matrix. The prominent example of a classically integrable theory which retains integrability at the quantum level is the principal chiral model (PCM) \cite{Luscher:1977uq,Goldschmidt:1980wq,Ogievetsky:1987vv}. Other classically integrable theories which are believed to be integrable at the quantum level include the $\l$-deformed model \cite{Sfetsos:2013wia}, as well as the more general integrable $\l$-models based on products of semi-simple groups \cite{Georgiou:2016urf,Georgiou:2017jfi,Georgiou:2018hpd,Georgiou:2018gpe}. 
These more general  models exhibit an interesting and rich renormalisation group  behaviour since they flow from a sum of WZW models in the UV to certain 2-dimensional conformal field theories  in the IR (see also \cite{Georgiou:2020eoo}).

Actually, one could, by imposing to the S-matrix the absence of particle production and factorisation, constrain the couplings and consequently the space of two-dimensional integrable theories. However, one faces an insurmountable obstacle. In order to be able to use standard field theoretical methods one should expand the action around a vacuum solution of the equations of motion. The most obvious such solution is the trivial vacuum \cite{Hoare:2018jim}. But in this case the excitations around 
this vacuum are all massless. It is unfortunate and quite surprising that when expanding around the "massless" trivial vacuum the connection between integrability and S-matrix is lost, namely in well-known 2-dimensional integrable models the tree-level massless S-matrix  exhibits particle production and fails to factorise \cite{Nappi:1979ig,Figueirido:1988ct,Hoare:2018jim,Horta}. Unfortunate because
the absence of particle production can not be used as a criterion to determine which target space geometries may correspond to integrable models \cite{Wulff:2017vhv}.
And surprising because the proof of classical integrability, which is based on the existence of a Lax connection,  makes no reference to whether the excitations of the theory are massless or massive.
On the other hand, it is true that the proof that  integrability implies factorisation and the absence of particle production has  an essential assumption. The excitations that scatter should be massive \cite{Parke:1980ki,Parke:1980ki-1,Arefeva:1974bk}.

As was vividly described in \cite{Hoare:2018jim}, the failure of the S-matrix to comply with the requirements of integrability may be due to the fact that many massless tree-level amplitudes suffer from IR ambiguities.\footnote{In the case of theories which can have amplitudes that are free from IR ambiguities the connection between integrability and the factorisation of the S-matrix seems to be valid, see \cite{Dubovsky:2012sh,Cooper:2014noa}. }
 Indeed, scattering amplitudes of massless excitations  may have $0/0$ ambiguities due to the fact that vertices and internal propagators may vanish simultaneously when the external momenta are put on-shell. 
In order to resolve these IR ambiguities one should regularise the amplitudes in some way. However, all regularisation prescriptions used so far in the literature are not consistent with integrability, they break it, leading to particle production \cite{Hoare:2018jim}. 

In this work, we propose a definition of the S-matrix of integrable $\s$-models with massless excitations which reestablishes the link between integrability and factorisation of the S-matrix. This is accomplished by expanding  the action of the theory around a non-trivial vacuum depending on a single parameter $\omega$, such that when one sends  $\omega \rightarrow 0$ one obtains the expansion around the trivial vacuum. One should use the Feynman rules derived from this Lagrangian in order to calculate the amplitudes $A(i \rightarrow f,\omega)$ and {\it only at the  end of the calculation} one should  set $\omega=0$ in order to recover the massless theory. 
Let us stress that  although the Lorentz symmetry is broken in the intermediate steps of the calculation it is restored in the final result for the S-matrix (see, for example, \eqref{M1pfin}).\footnote{ Previous studies of massless S-matrices for integrable theories in relation to finite-density TBA  can be found in \cite{Zamolodchikov:1992zr,Fendley}.}  At this point, let us mention that, although we believe that this statement is always true, we do not have a general proof at our disposal.  Finally, let us mention that our definition can be applied to the large class of $\s$-models whose target space geometry possesses one or more isometries.

We exemplify our method by considering the important example of the isotropic PCM with group SU(2).

\section{Definition of the S-matrix: regularisation of the IR ambiguities}\label{DEF}

The question that we address in this section is the following. Is it possible to define the S-matrix of an integrable $\sigma$-model with massless excitations in a way consistent with integrability  or this is inherently impossible and integrability is necessarily anomalous?

In what follows, we will argue that it is, indeed, possible to define a massless S-matrix which exhibits no particle production, admits only equal sets of initial and final momenta and factorises. 
We will consider the large class  of integrable $\sigma$-models with one isometry at least.
Our prescription for the evaluation of the scattering amplitudes is as follows:
\begin{mdframed}[backgroundcolor=blue!5, leftmargin=0cm, rightmargin=0cm, topline=false,
	bottomline=false, leftline=false, rightline=false] 
\vspace{-0.0cm}

\begin{itemize}

\item[\ding{182}] 
The key point is to expand the Lagrangian around the non-trivial vacuum 
\be\label{nt-vac}
 X^a=X^a_0={\rm const.},\,\,\,\,\,     \,\,\,\,\, \, x^i=\omega^i \tau,  \,\,\,\,\, {\rm with} \,\,\,\,\, \omega^i=\omega\,\, {\rm or}\,\, 0,
\ee
depending on how many isometries we turn on.
Here $x^i$ denotes the different isometries of the theory while $X^a$ are the non-isometric coordinates. One should employ as many isometries one needs in order to make massive, at least, the fields associated to the non-isometric directions.
We, then, use the Feynman rules derived from this Lagrangian in order to calculate the amplitudes $A(i \rightarrow f,\omega)$ and {\it only at the  end of the calculation} we set $\omega=0$ in order to recover the massless theory, namely
\be\label{def}
A(i \rightarrow f)=\lim_{\omega\rightarrow 0} A(i \rightarrow f,\omega)\, .
\ee
The diagrams whose IR divergences  are not regulated by expanding around the non-trivial vacuum will be regulated by keeping the $i \epsilon$ term in the propagator.
\item[\ding{183}] For all amplitudes which are free from IR ambiguities, the above prescription gives the same result as the one obtained by using the Feynman rules derived from the Lagrangian expanded around the trivial vacuum
\be\label{t-vac}
 X^a=X^a_0={\rm const.},\,\,\,\,\,     \,\,\,\,\, \, x^i=0 \,\,\,\,\, \forall \,\,i.
\ee
 
\end{itemize}
\end{mdframed}

To exemplify the points above let us consider two examples,   the  PCM based on the group $SU(2)$ and  the  coset model with target space $S^5\cong SO(6)/SO(5) $. The case of the PCM is treated in great detail in section \ref{PCM}.  For  the coset model the corresponding target space metric reads
\begin{equation}\label{metrics}
ds_{S^5}^2= R^2\big(d\theta^2+\cos^2\theta \,d\phi_3^2+\sin^2\theta\,\, d\Omega_3^2\big),\,
d\Omega_3^2=d\psi^2+\cos^2\psi \,d\phi_1^2+\sin^2\psi\, d\phi_2^2.
 \end{equation}
We now make the following change of variables 
\be
\theta=\frac{\tilde\theta}{R}, \qq \phi_3=\omega \tau+\frac{\varphi}{R}. 
\ee
and expand for large values of the radius $R$, order by order.  
In these coordinates $\varphi$ remains massless while all the other four coordinates $\tilde \theta, \psi, \phi_1,\phi_2$, or rather their Euclidean counterparts $x_i, i=1,..,4$ defined by the relation $d\tilde\theta^2+\tilde\theta^2 \,d\Omega_3^2=dx_1^2+dx_2^2+dx_3^2+dx_4^2, \, \tilde\theta^2 =\sum_{i=1}^4 x_i^2$
, acquire mass. At this point, let us mention that in more complicated models one may need to turn on more than one isometries in order to give masses to all the excitations which originate from the non-isometric directions.  As mentioned in the introduction, the existence of isometries is crucial for our definition to work. Thus, by turning on 
a sufficient number of isometries all non-isometric fields acquire mass. This is necessary for our prescription to work and as argued below \eqref{actexpa} it can be  generically achieved.

Let us now be a bit more general. We consider a generic $\sigma$-model describing the evolution of $N=N_1+N_2$  degrees of freedom $z^\mu=\{x^i, X^a \},\, \mu=1,...,N$, with $N_1$ being the  isometric directions denoted by $x^i,\,\,i=1,...,N_1$ and $N_2$ being the non-isometric ones  $X^a,\,\,a=1,...,N_2$. Thus, we have
\begin{equation}
\label{action}
S= \frac{1}{2\pi}\int d^2\sigma\,  \cL\ ,\qq
 \cL = (G_{\m\n}+ B_{\m\n}) \del_+z^\m\del_-z^\n\,,
\end{equation}
with $G_{\m\n}$ and $B_{\m\n}$ being  the metric and the  antisymmetric tensor fields, respectively. 
The relation between the world-sheet coordinates $\s^\pm$ and $(\tau,\s)$ are given below
\be
\s^\pm=\tau\pm\s\,,\quad \del_\pm=\frac12\left(\del_\tau\pm\del_\s\right)\,,\quad \text{d}^2\s=d\tau\,d\s\,.
\ee
The action can thus be rewritten as
\begin{eqnarray}
\label{actga}
&&S =\frac{1}{8 \pi} \int \text{d}^2\s \Big( \eta^{\alpha \beta} \partial_\alpha z^\m \partial_\beta z^\n G_{\m\n}(X^a) - \bar \e^{\alpha \beta} \partial_\alpha z^\m \partial_\beta z^\n B_{\m\n} (X^a)
  \Big)=\\
 &&  \frac{1}{8 \pi} \int \text{d}^2\s \Big(  \partial_\alpha X^a \partial^\alpha X^b G_{ab}(X^a)+\partial_\alpha x^i \partial^\alpha x^j G_{ij}(X^a) +2\, \partial_\alpha x^i \partial^\alpha X^a G_{ia}(X^a)\nonumber  \\
&& - \bar \e^{\alpha \beta} \partial_\alpha X^a \partial_\beta X^b B_{ab} (X^a)-\bar \e^{\alpha \beta} \partial_\alpha x^i \partial_\beta x^j B_{ij} (X^a)-2\,\bar \e^{\alpha \beta} \partial_\alpha x^i \partial_\beta X^a B_{ia} (X^a)
  \Big)\nonumber \ ,
\end{eqnarray}
where our conventions are: $\eta_{\alpha \beta}=\diag(1,-1)$ with $\alpha, \beta=0,1$ and $\bar \e^{01}=1=-\e^{10}$.
Next we expand around the non-trivial solution $X^a=X^a_0={\rm const.}$ and $x^i=\omega^i \tau$, as follows
\be\label{expa}
X^a=X^a_0+{\tilde X^a \over R},\qquad x^i=\omega^i \tau+{\tilde x^i \over R}\, .
\ee
Here $R$ is some scale of the geometry, i.e. $G_{\m\n}\sim R^2$ and $B_{\m\n}\sim R^2$ , which will play the role of the inverse coupling constant (see for example \eqref{metrics}). 

Before proceeding, let us comment under which conditions the vacuum \eqref{nt-vac} is a solution of the equations of motion (eom). 
 The eom for $X^a$ are solved for the general ansatz  $X^a=X^a_0,\, x^i=\omega^i \tau$ (corresponding to \eqref{expa}) as long as the set of the $N_2$ equations 
\be\label{eoms}
 \sum_{ij} \omega^i \omega^j{\partial G_{ij} \over \partial X^a}=0
\ee
admits constant solutions, i.e. $X^a(\tau,\sigma)=X^a_0$. 
Once the relations \eqref{eoms}
can be solved for constant values  of the $X^a$  the eom for the isometric coordinates are also automatically satisfied. If one makes a particular choice for $\omega^i$, like the one made in \eqref{nt-vac}, then \eqref{eoms} reduces to simpler equation involving a single $\omega$, namely $\omega^2 \sum^{'}_{ij}  {\partial G_{ij} \over \partial X^a}=0$, where the prime in the sum means that the sum  is restricted to the isometries $i$ and $j$ for which the corresponding $\omega^i$ are non-zero.

Plugging now \eqref{expa} in \eqref{actga} one obtains for the terms which are up to order $R^0$ in the large $R$ expansion the following expression
\begin{eqnarray}
\label{actexpa}
 && S = \frac{1}{8 \pi R^2} \int \text{d}^2\s \Big(  \partial_\alpha \tilde X^a \, \partial^\alpha \tilde X^b \,G_{ab}(X_0^a)+\partial_\alpha \tilde x^i \, \partial^\alpha \tilde x^j \,G_{ij}(X_0^a)+ \omega^i \omega^j  \tilde X^a \tilde X^b {\partial^2 G_{ij}(X_0^a)\over \partial X^a  \, \partial X^b} \nonumber  \\
&&+2\, \omega^i \partial_\tau \tilde x^j \tilde X^a {\partial G_{ij}(X_0^a)\over \partial X^a }+2\, \omega^i \partial_\tau \tilde X^a \tilde X^b {\partial G_{ia}(X_0^a)\over \partial X^b }
\nonumber  -2\, \omega^i \partial_\sigma \tilde x^j \tilde X^a {\partial B_{ij}(X_0^a)\over \partial X^a }\\
&& -2\, \omega^i \partial_\sigma \tilde X^a \tilde X^b {\partial B_{ia}(X_0^a)\over \partial X^b }+
\mathcal O ({1 \over R})
  \Big) \ .
\end{eqnarray}
To derive \eqref{actexpa} we have used the condition  \eqref{eoms}, we have dropped total derivative terms and we have ignored the constant term 
$\omega^i \omega^jG_{ij}(X_0^a)$. Furthermore, terms of  the form 
$(\partial_\tau \tilde x^i\partial_\sigma \tilde X^a-\partial_\sigma \tilde x^i\partial_\tau \tilde X^a)B_{ia}(X_0^a)$ vanish after partially integrating twice the first term in the parenthesis. Notice the third term in the first line of \eqref{actexpa} which is quadratic in the $\tilde X^a$ coordinates and is proportional to the mass matrix in the subspace of the non-isometric coordinates. Notice that, as long as the determinant of the matrix 
$\omega^i \omega^j  {\partial^2 G_{ij}(X_0^a)\over \partial X^a  \, \partial X^b}$ is non-zero, all the non-isometric directions acquire mass.

Two important comments are in order. Firstly, notice that by expanding around the non-trivial vacuum \eqref{nt-vac} even the quadratic (free) part of the action, see \eqref{actexpa}, is not Lorentz invariant since there are terms involving a single $\tau$ or $\sigma$ derivative. This directly implies that, in the general case, the dispersion relations derived from the diagonalisation of \eqref{actexpa} will neither  be Lorentz invariant. In addition the non-diagonal terms of the second and third line in \eqref{actexpa}
will generically generate a mixing between the isometric and non-isometric coordinates. It is in the diagonal basis in which one should calculate the scattering amplitude before taking the massless limit 
\eqref{def}. Secondly, one may have an objection regarding the prescription \eqref{def} since one may use the {\it classical} scale invariance of the $\sigma$-model in order to make $\omega$ disappear from the action.  Indeed, by rescaling appropriately $\tau$ and $\sigma$ one can set $\omega=1$.
However, we are interested in the massless limit which is obtained when $\omega=0$ and this can never be reached once $\omega$ is eliminated from the action. Thus, we explicitly keep the parameter $\omega$ in the action and reach the massless limit by sending $\omega\rightarrow 0$.

Let us now comment on the logic behind the definition \eqref{def}. 
To illustrate the main point, consider the simple case of an integrable model with no B-field and with the metric being such that the first and second term appearing in the second line of \eqref{actexpa} being absent. An explicit example  of this kind is the model studied in section \ref{PCM}. By expanding around the non-trivial vacuum $x^i=\omega^i \tau$ it is generically possible to make all particles, except those related to the non-trivial isometries $x^i$, massive\footnote{Although this statement is generically true, there are exceptional cases in which the particles of the theory do not acquire mass even when expanding around the aforementioned non-trivial vacuum. Such an example is the model considered in \cite{Nappi:1979ig}.} with their mass being proportional to the parameter $\omega$, i.e. $m_a \sim \omega$. Indeed the third term in the first line of \eqref{actexpa} which is quadratic in the $\tilde X^a$ coordinates plays the role of  the mass matrix\footnote{Here we do not examine issues regarding whether the mass matrix is positive definite or not.} in the subspace of the non-isometric coordinates. The excitations related to the isometries around which we expand remain, however, massless.
 But this is harmless since the massless excitations appear only quadratically in the action since they originate from the isometries.
At any instance, these fields can be integrated out from the action so that one ends up with the massive excitations only. This is the reason why one should expect that using this procedure the connection between the S-matrix and integrability will be restored even after taking the massless limit $\omega \rightarrow 0$ which implies that all particles become again massless $m_a\sim \omega \rightarrow  0$.

A few more comments are in order. The first concerns the limit appearing in \eqref{def}. As we will see in explicit examples later on, individual Feynman diagrams may diverge in the limit $\omega \rightarrow  0$, namely they may behave like $\frac{1}{\omega^{2n}},\, n \in \mathbb N^*$. However, these divergences cancel when all diagrams contributing in a certain amplitude are summed  giving a well-defined finite result in the massless limit. 
At this point let us mention that there are amplitudes for which each single diagram is finite in the massless limit. The second comment concerns the non-trivial vacuum $x_i=\omega^i \tau$ with 
$\omega^i=\omega \,\,{\rm or} \,\,0$ around which we expand in order to give temporarily masses to the particles. This vacuum is a solution of the complete set of equations of motion of the theory. 
One may think of expanding around a more generic solution of the type $x^i=\omega^i \tau$ which introduces more than one parameters $\omega^i$. However, it is not obvious to the present author, if the final result for the amplitude will depend or not on ratios of the parameters $\frac{\omega^i}{\omega^j}$ when taking the massless limits $\omega^i \rightarrow  0$ and $\omega^j\rightarrow  0$. 
In this  case the physical interpretation of the result is obscured.  We, thus, evaluate the amplitudes by expanding around the single parameter  vacuum solution $x^i=\omega \tau$. A related issue is the following.  Suppose that there is more than one solution of the set of equations \eqref{eoms}. Although we have not checked explicitly, we think that the S-matrix will exhibit the properties of factorisation and lack of particle production for each of those solutions.

Finally and most importantly, let us stress that the aforementioned way of regulating the IR problem is conceptually very different from those used so far in the literature. This is so because our prescription introduces new vertices in the theory. It is precisely their contribution which cancels the non-zero result in the $2 \rightarrow 4$ amplitude, for example. This is radically different compared to the regularisations used so far, namely the $i \epsilon$ regularisation in the propagators or the one in which one artificially gives  the same mass to all particles, internal and external \cite{Hoare:2018jim}. At this point, let us mention that our way of regularising the IR ambiguities is natural since the mass of each excitation is implemented not by hand but in a natural and dynamical way by expanding around the non-trivial vacuum \eqref{nt-vac}.
An ultimate comment is the following. Although the Lorentz symmetry  is broken in the intermediate steps of the calculation it is restored in the final result for the S-matrix (see, for example, \eqref{M1pfin}).

\section{The case of the $SU(2)$ PCM}\label{PCM}

To exemplify the general discussion of the last section, we focus on the case of the $SU(2)$ principal chiral model.  In the parametrisation \eqref{group-element}, this theory has an apparent isometry which we denote by $\phi$.
By parametrising the $SU(2)$ group element as 
\be\label{group-element}
g=\left(
\begin{array}{cc}
 \cos a+i \sin a \cos b & e^{-i \phi} \sin a \sin b \\
 -e^{i \phi} \sin a \sin b & \cos a-i \sin a \cos b \\
\end{array}
\right)
\ee
the  Lagrangian of the PCM takes the following form
\begin{eqnarray}\label{Lag}
{\cal L} = \frac{R^2}{2}  \Big( \partial_{\mu}a \,  \partial^{\mu}a + \sin^2a\,(\partial_{\mu}b \,  \partial^{\mu}b + \sin^2b \,\partial_{\mu}\phi \,  \partial^{\mu}\phi)
 \Big)
\end{eqnarray}
where $R$ is the radius of the space. Furthermore, our conventions for the world-sheet metric is the following: $\eta_{\mu \nu}=\diag(1,-1)$ with $\mu,\nu=0,1$.
The next step is to make  the following change of variables
\be\label{changeofvar}
a=\frac{\pi}{2}+\frac{\alpha}{R}, \qquad  b=\frac{\pi}{2}+\frac{\beta}{R}, \qquad \phi=\omega \tau+\frac{\gamma}{R}
\ee
and subsequently expand the Lagrangian for large values of the radius $R$  keeping as many terms one wishes. In what follows we will need to keep terms up to order $\frac{1}{R^4}$.

In this way one obtains the Lorentz violating Lagrangian
\be\label{Lag-exp}
{\cal L}^{(\omega)}={\cal L}_0+\frac{{\cal L}_1}{R}+\frac{{\cal L}_2}{R^2}+\frac{{\cal L}_3}{R^3}+\frac{{\cal L}_4}{R^4} +{\cal O}(\frac{1}{R^5})\, ,
\ee
where 
\be\label{Lag0}
{\cal L}_0=\frac{1}{2}\Big( \partial_{\mu}\alpha   \partial^{\mu}\alpha + \partial_{\mu}\beta  \partial^{\mu}\beta +\partial_{\mu}\gamma   \partial^{\mu}\gamma-\omega^2 (\alpha^2+\beta^2)\Big).
\ee
As mentioned above, the excitations $\alpha$ and $\beta$ acquire the mass, which here happens to be the same, while the excitation originating from the isometry, $\gamma$, remains massless.
The interaction terms take the form
\be\label{Lag1}
{\cal L}_1=-\omega(\alpha^2+\beta^2)\partial_{\tau} \gamma,
\ee
\be\label{Lag2}
{\cal L}_2=\frac{\omega^2}{6}(\alpha^4+\beta^4)-\frac{1}{2}(\alpha^2+\beta^2)\partial_{\mu}\gamma   \partial^{\mu}\gamma-\frac{1}{2}\alpha^2\partial_{\mu}\beta   \partial^{\mu}\beta
+\frac{\omega^2}{2}\alpha^2 \beta^2,
\ee
\be\label{Lag3}
{\cal L}_3=\frac{\omega}{3}(\alpha^4+\beta^4+3\alpha^2 \beta^2)\partial_{\tau} \gamma.
\ee
Finally, 
\be\label{Lag4}
{\cal L}_4=-\frac{\omega^2}{45}(\alpha^6+\beta^6)-\frac{\omega^2}{6}(\alpha^2 \beta^4+\alpha^4 \beta^2)
+\frac{1}{6}(3\,\alpha^2 \beta^2+\alpha^4+\beta^4)\partial_{\mu}\gamma \,  \partial^{\mu}\gamma+\frac{1}{6}\alpha^4 \partial_{\mu}\beta \,  \partial^{\mu}\beta\,.
\ee
At this point notice that, although the vertices proportional to powers of $\omega$ vanish in the limit $\omega \rightarrow 0$, they may give finite contributions to the amplitudes when multiplied with a propagator which becomes on-shell. It is precisely these contributions which cancel the non-zero result 
in the amplitudes which exhibit particle production.
Finally, the Lagrangian expanded around the trivial vacuum $g=\left(
\begin{array}{cc}
\, \,\,\,0 & 1 \\
 -1 & 0 \\
\end{array}
\right)$, that is $a={\pi \over 2}$,  $b={\pi \over 2}$ and $\phi=0$, takes the simpler form
\begin{eqnarray}\label{Lagom0}
{\cal L}^{(\omega=0)}=&&
\frac{1}{2}\Big( \partial_{\mu}\alpha   \partial^{\mu}\alpha + \partial_{\mu}\beta  \partial^{\mu}\beta +\partial_{\mu}\gamma   \partial^{\mu}\gamma-\frac{1}{R^2}(\alpha^2+\beta^2)\partial_{\mu}\gamma   \partial^{\mu}\gamma-\frac{1}{R^2}\alpha^2\partial_{\mu}\beta   \partial^{\mu}\beta\Big)\nonumber \\
+&&\frac{1}{6 R^4}\Big((3\,\alpha^2 \beta^2+\alpha^4+\beta^4)\partial_{\mu}\gamma \,  \partial^{\mu}\gamma+\alpha^4 \partial_{\mu}\beta \,  \partial^{\mu}\beta\Big)\,.
\end{eqnarray}
Notice that in \eqref{Lagom0} all excitations are massless.

\section{The $2\rightarrow 2$ S-matrix for the $SU(2)$ PCM}\label{2to2}

In this section we present the $2\rightarrow 2$ scattering amplitudes. It is straightforward to calculate them, so we just present the final results.

$ \bullet\,\,\, \a_1+\b_2\rightarrow \a_3+\b_4$ scattering amplitude \\
According to the highlighted points of section \ref{DEF} the allowed kinematics of the particles will be determined as the $\omega \rightarrow 0 $ limit of the energy and momentum conservation laws of the massive theory, namely taking into account that $E=\sqrt{p^2+m^2}=\sqrt{p^2+\omega^2}$ we have 
\be \label{kin1}
\left.  \begin{array}{cc} E_1+E_2=E_3+E_4\\
           p_1+p_2=p_3+p_4 \end{array}   \right \}
\Longrightarrow \left.  \begin{array}{cc} p_1=p_3\\
           p_2=p_4 \end{array}   \right \} \qq \text{or} \left. \qq \begin{array}{cc} p_1=p_4\\
           p_2=p_3 \end{array}   \right \}
\ee
In what follows and in accordance to the second highlighted point in section \ref{DEF}, we will assume that these are the only solutions of the energy-momentum conservation laws even at the massless limit $\omega=0$. The first solution in \eqref{kin1} corresponds to the process in which the outgoing $\a$ particle has the same momentum as the incoming particle of the same species while in the second solution the particles $\a$ and $\b$ exchange momenta. In the massless limit the particles have the following 2-momenta $p_i^\mu=(|p_i|,p_i),\,\,i=1,\dots, 4$. We will denote the first process by $A_{\a\b\rightarrow\a\b}$ while the second by $A_{\a\b\rightarrow\b\a}$. The result for the amplitudes is easily found to be
\be\label{ab>ab}
A_{\a\b\rightarrow\a\b}=0\qquad A_{\b\a\rightarrow\a\b}=A_{\a\b\rightarrow\b\a}=-2\, i\,\l \, p_2\cdot p_4=-2\, i \, \l \, p_1^\mu \,p_{2 \mu}=-2\, i\, \l \, p_1\cdot p_2,
\ee\\
where $\l={1 \over R^2}$.
Next we calculate the amplitudes involving all possible combinations of species.\\

$\bullet \,\,\,\a_1+\a_2\rightarrow \b_3+\b_4$ and $\b_1+\b_2\rightarrow \a_3+\a_4$ scattering amplitudes
\be\label{aa>bb}
A_{\a\a\rightarrow\b\b}=A_{\b\b\rightarrow\a\a}=2\, i\, \l \,p_1\cdot p_2
\ee

$\bullet \,\,\,\a_1+\a_2\rightarrow \a_3+\a_4$ and $\b_1+\b_2\rightarrow \b_3+\b_4$ scattering amplitudes
\be\label{aa>aa}
A_{\a\a\rightarrow\a\a}=0\qquad A_{\b\b\rightarrow\b\b}=0
\ee

$\bullet \,\,\,\a_1+\g_2\rightarrow \a_3+\g_4$  scattering amplitudes
\be\label{ag>ag}
A_{\a\g\rightarrow\a\g}=0\qquad A_{\a\g\rightarrow\g\a}=-2\, i\,\l \, p_2\cdot p_4=-2\, i\, \l \,p_1\cdot p_2
\ee

$\bullet \,\,\,\b_1+\g_2\rightarrow \b_3+\g_4$  scattering amplitudes
\be\label{ag>ag}
A_{\b\g\rightarrow\b\g}=0\qquad A_{\b\g\rightarrow\g\b}=-2\, i\,\l \, p_2\cdot p_4=-2\, i\, \l \,p_1\cdot p_2
\ee

$\bullet\,\,\,\a_1+\a_2\rightarrow \g_3+\g_4$ , $\b_1+\b_2\rightarrow \g_3+\g_4$, $\g_1+\g_2\rightarrow \a_3+\a_4$ and  $\g_1+\g_2\rightarrow \b_3+\b_4$ scattering amplitudes
\be\label{ag>ag}
A_{\a\a\rightarrow\g\g}=A_{\g\g\rightarrow\a\a}=2\, i\, \l \,p_1\cdot p_2\qquad A_{\b\b\rightarrow\g\g}=A_{\g\g\rightarrow\b\b}=2\, i\, \l \,p_1\cdot p_2
\ee
Notice that the $2 \rightarrow 2$ amplitudes can be written in a compact form ($\epsilon_{123}=1$)
\be\label{compactA}
A\big(X^a(p_1)X^b(p_2)\rightarrow X^c(p_1) X^d(p_2)\big)=2\, i\, \l \,\epsilon_{ack} \epsilon_{bdk}\,p_1\cdot   p_2, 
\ee
after the identification $X^1\equiv \a,\, X^2\equiv \b,\, X^3\equiv \g$.
This is, of course, reminiscent of the underlying $SU(2)$ symmetry of the theory.

\section{No particle production for the $SU(2)$ PCM}

\subsection{A $2\rightarrow 4$ amplitude}
In order to make clear how our prescription works we firstly consider an amplitude in which two incoming particles scatter to produce four outgoing particles. More precisely, we consider the particle production process in which two particles of type $\a$ scatter to produce four particles of type $\b$, namely
\be\label{a+a+->b-b-b+b+}
\a(p_1^+)\,  \a(p_2^-)\rightarrow \b(p_3^-)\,\b(p_4^-)\,\b(p_5^+)\, \b(p_6^+),
\ee
where the plus and minus associated to each momentum denotes if the particle is right or left moving respectively. Thus, we have chosen the process for which 
\be\label{2to4-kin}
p_i^\mu=p_i^+=(p_i, p_i),\,i=1,5,6\qquad   {\rm and} \qquad p_j^\mu=p_2^-=(p_j, -p_j),\,j=2,3,4\,
\ee 
where all the energies of the particles are taken positive, i.e. $p_i> 0$ and $p_j> 0$. 
This amplitude should be zero in an integrable theory. However, one can easily check that by using the 
Lagrangian \eqref{Lagom0} (see also \cite{Hoare:2018jim}) this amplitude has a non-zero value. In the rest of this section, we will show that by using the definition described in section \ref{DEF} the amplitude does vanish.

In the massless limit, energy and momentum conservation implies that $p_1=p_5+p_6$ and $p_2=p_3+p_4$. But here we will impose the energy-momentum conservation with the particles $\a$ and $\b$ being slightly massive in which case the conservation laws imply
\begin{eqnarray}\label{con-laws1}
&&p_1=p_5+p_6+a,\, p_2=p_3+p_4+a,\\ 
&&a={1\over 16} m^2 \Big(m^2 \big(\sum_{i=1}^2{1\over p_i^3} - \sum_{i=3}^6{1\over p_i^3}\big) + 
   4 \big(-\sum_{i=1}^2{1\over p_i} + \sum_{i=3}^6{1\over p_i}\big)\Big)+\mathcal O(m^6)\nonumber\, .
\end{eqnarray}
In \eqref{con-laws1} and in all similar equations, $p_i$ denotes the space components of the 2-momenta
 $p_i^\mu$, except in the occasions where a dot (inner product) appears. In the latter case $p_i$ will denote the 2-momentum of the corresponding particle. Finally, let us note that  the propagators for the massive excitations $\a$ and $\b$ is given by 
\be
D(p)={i \over p^2-m^2+i \epsilon}
\ee
while for the massless field $\g$ is given by 
\be\label{massless-prop}
\tilde D(p)={i \over p^2+i \epsilon}
\ee
with $m=\omega$ and $p^2=p_\mu p^\mu$.

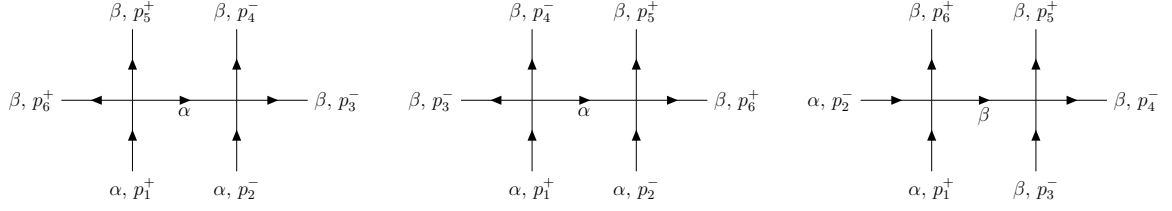
\begin{figure}

\resizebox{5cm}{!}{
\begin{tikzpicture}

\begin{feynman}
\vertex (a);
\vertex [below=of a](i1) {\(\alpha\), \( p_1^+\)};
\vertex [above=of a](f3) {\(\b\), \( p_5^+\)};
\vertex [left=of a](f4) {\(\b\), \( p_6^+\)};
\vertex [right=2.2cm of a](b);
\vertex [above=of b](f2){\(\b\), \( p_4^-\)};
\vertex [right=of b](f1){\(\b\), \( p_3^-\)};
\vertex [below=of b](i2){\(\alpha\), \( p_2^-\)};

 \diagram* {
(i1) -- [fermion] (a)-- [fermion, edge label'=\(\a\)] (b)-- [fermion] (f2),
(b)-- [fermion] (f1),
(i2) -- [fermion] (b),
(a) -- [fermion] (f3),
(a) -- [fermion] (f4),
}; 

\hspace{240pt}

\vertex (a);
\vertex [below=of a](i1) {\(\alpha\), \( p_1^+\)};
\vertex [above=of a](f3) {\(\b\), \( p_4^-\)};
\vertex [left=of a](f4) {\(\b\), \( p_3^-\)};
\vertex [right=2.2cm of a](b);
\vertex [above=of b](f2){\(\b\), \( p_5^+\)};
\vertex [right=of b](f1){\(\b\), \( p_6^+\)};
\vertex [below=of b](i2){\(\alpha\), \( p_2^-\)};

 \diagram* {
(i1) -- [fermion] (a)-- [fermion, edge label'=\(\a\)] (b)-- [fermion] (f2),
(b)-- [fermion] (f1),
(i2) -- [fermion] (b),
(a) -- [fermion] (f3),
(a) -- [fermion] (f4),
}; 
 
\hspace{240pt}

\vertex (a);
\vertex [below=of a](i1) {\(\alpha\), \( p_1^+\)};
\vertex [above=of a](f3) {\(\b\), \( p_6^+\)};
\vertex [left=of a](f4) {\(\a\), \( p_2^-\)};
\vertex [right=2.2cm of a](b);
\vertex [above=of b](f2){\(\b\), \( p_5^+\)};
\vertex [right=of b](f1){\(\b\), \( p_4^-\)};
\vertex [below=of b](i2){\(\b\), \( p_3^-\)};

 \diagram* {
(i1) -- [fermion] (a)-- [fermion, edge label'=\(\b\)] (b)-- [fermion] (f2),
(b)-- [fermion] (f1),
(i2) -- [fermion] (b),
(a) -- [fermion] (f3),
(f4) -- [fermion] (a),
};

\end{feynman}

\end{tikzpicture}
}

\caption{Diagrams with a zero contribution in the limit $\omega\rightarrow 0$. There are 3 additional diagrams, similar to the rightmost one, which are obtained by cyclic  permutations of the $\b$ external particles. They are zero, too. }
\label{zero}
\end{figure}

Before proceeding let us comment on the scaling of the different inner products. The inner product of two momenta which in the massless limit, $\omega\rightarrow 0$, represent a left-moving and a right-moving particle is finite and is receiving correction proportional to $m^2=\omega^2$. On the other hand, the inner product of two momenta which in the massless limit, $\omega\rightarrow 0$, represent
two left-moving or two right-moving particles is proportional to $m^2=\omega^2$ and tends to zero in the massless limit, as it should. Similarly, the propagator whose momentum is purely left or right moving behaves as $D(p)\sim {1 \over m^2}+{\mathcal O}(m^0)$.
\subsubsection{Diagrams of type I }
These diagrams involve two 4-point vertices each of which can be either 
$-{1 \over 2}\a^2\partial_\mu\b \partial^\mu\b$ or ${\omega^2 \over 2}\a^2\b^2$.
Given the comments below \eqref{massless-prop} it straightforward to see that ambiguous diagrams of figure \ref{zero} give vanishing contributions. The left and middle diagrams are proportional to 
$(m^2)^2{1\over m^2}\sim m^2\rightarrow 0$. When both vertices are of the type 
$-{1 \over 2}\a^2\partial_\mu\b \partial^\mu\b$, the two powers of $m^2$ in the numerator come from the two inner products of the $\b$ particles, i.e. $p_3\cdot p_4$ and $p_5\cdot p_6$, while the $m^2$ in the denominator from the $\a$ particle propagator which goes on-shell in the massless limit. 
If one or both vertices are of the type ${\omega^2 \over 2}\a^2\b^2$ the scaling with respect to $m$ is the same since the $m^2=\omega^2$ which previously was coming from the inner products now comes from the the fact that the vertex/vertices is proportional to $\omega^2$.
The right diagram of figure \ref{zero} is zero because now the $\a$ propagator is finite in the massless limit and there is only a factor of 
$\omega^2\rightarrow 0$ originating from the right vertex which involves four $\b$ excitations (see the first term in the right hand side of \eqref{Lag2}).

\begin{figure} 
\centering

\resizebox{8cm}{!}{
\begin{tikzpicture}


\begin{feynman}
\vertex (a);
\vertex [below=of a](i1) {\(\alpha\), \( p_1^+\)};
\vertex [above=of a](f3) {\(\b\), \( p_4^-\)};
\vertex [left=of a](f4) {\(\b\), \( p_6^+\)};
\vertex [right=2.5cm of a](b);
\vertex [above=of b](f2){\(\b\), \( p_5^+\)};
\vertex [right=of b](f1){\(\b\), \( p_3^-\)};
\vertex [below=of b](i2){\(\alpha\), \( p_2^-\)};

 \diagram* {
(i1) -- [fermion] (a)-- [fermion, edge label'=\(\a\)] (b)-- [fermion] (f2),
(b)-- [fermion] (f1),
(i2) -- [fermion] (b),
(a) -- [fermion] (f3),
(a) -- [fermion] (f4),
}; 

\end{feynman}

\end{tikzpicture}
}
\caption{Diagram of type I contributing to the amplitude \eqref{a+a+->b-b-b+b+} in the limit $\omega\rightarrow 0$. One should add 3 more diagrams. One  with $p_3\leftrightarrow p_4$,  one with $p_5\leftrightarrow p_6$ and one with $p_3\leftrightarrow p_4 \land p_5\leftrightarrow p_6$ exchanged.}
 \label{A0f}

\end{figure}
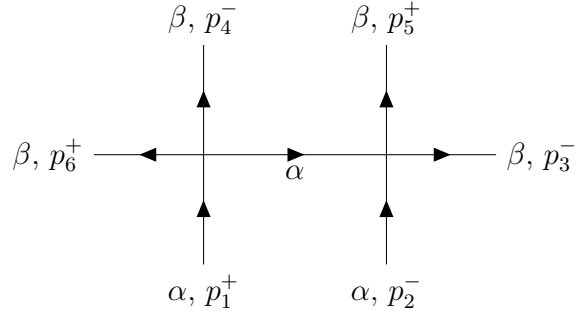

The contribution of the diagrams in figure \ref{A0f} is given by \footnote{The factor of $2^4$ in \eqref{A0} comes from the combinatorics of the diagram, the factor of $(-i)^4$ from the 4 derivatives acting on the $\b$ particles while the factor of $\Big(-{i \l\over 2 }\Big)^2 $ from the coupling constants of the two vertices.}
\begin{eqnarray}\label{A0}
A_0&=&2^4\Big(-{i \over 2 }\l\Big)^2  D(p_2-p_3-p_5)(-i)^4 p_3\cdot p_5\, p_4\cdot p_6+(p_3\leftrightarrow p_4)+(p_5\leftrightarrow p_6)+\nonumber \\ 
&&(p_3\leftrightarrow p_4 \land p_5\leftrightarrow p_6)\overset{\omega\rightarrow 0}{=}2 i \,\l^2\,p_1\cdot p_2=4 i \,\l^2\,p_1 p_2.
\end{eqnarray}
The meaning of the various parentheses is the following.  For example, the meaning of $(p_3\leftrightarrow p_4)$ is that one should add the term written explicitly (first term in \eqref{A0}) but with the momenta $p_3^\mu$ and $p_4^\mu$ exchanged.
At this point, let us mention that each of the two vertices in figure \ref{A0f} should be $-{1 \over 2}\a^2\partial_\mu\b \partial^\mu\b$ since the propagators are finite and the presence of a vertex of the type ${\omega^2 \over 2}\a^2\b^2$ would make this contribution vanish in the limit $\omega\rightarrow 0$.
$A_0$ in \eqref{A0} is the complete contribution to the scattering process \eqref{a+a+->b-b-b+b+} if one uses the Lagrangian \eqref{Lagom0} and is, of course, nonzero. In what follows we will see that this non-zero result will be canceled by the additional diagrams containing vertices that are present only when expanding around the nontrivial vacuum \eqref{nt-vac}. Their contribution is given below.

\begin{figure}

\resizebox{10cm}{!}{
\begin{tikzpicture} 
\begin{feynman}
\vertex (i1){\(\alpha\), \( p_1^+\)};
\vertex [right=2.7 of i1](a);
\vertex [right=6.0cm of a](b);
\vertex [right=of b](i2){\(\alpha\), \( p_2^-\)};
\vertex [above=2.2cm of a](c);
\vertex [above=2.2cm of b](d);
\vertex [above right=of d](f1) {\(\beta\), \(p_3^-\)};
\vertex [above left=of d](f2) {\(\beta\), \(p_4^-\)};
\vertex [above right=of c](f3) {\(\beta\), \(p_5^+\)};
\vertex [above left=of c](f4) {\(\beta\), \(p_6^+\)};

\diagram* {
(i1) -- [fermion] (a)--  [fermion, edge label'=\(\a\)] (b),
(i2) -- [fermion] (b),
(a)-- [red,boson, edge label'=\(\gamma\)] (c)-- [fermion] (f3),
(c) -- [fermion] (f4),
(b)-- [red,boson, edge label'=\(\gamma\)] (d)-- [fermion] (f1),
(d) -- [fermion] (f2),
}; 

\hspace{450pt}

\vertex (a);
\vertex [below left=of a](i1) {\(\alpha\), \( p_1^+\)};
\vertex [below right=of a](i2) {\(\alpha\), \( p_2^-\)};
\vertex [above right=2.7cm of a](b);
\vertex [above left=2.9cm of a](c);
\vertex [above left=of b](f3) {\( \beta\)}; 
\vertex [above right=of b](f4) {\(\beta\), \(p_3^-\)};
\vertex [above right=of c](f2) {\(\beta\), \(p_6^+\)}; 
\vertex [above left=of c](f1) {\(\beta\), \(p_5^+\)};

 \diagram* {
(i2) -- [fermion] (a)-- [red,boson, edge label'=\(\gamma\)] (b)-- [fermion] (f4),
(b)-- [fermion, momentum'={[arrow style=blue]\(p_4^-\)}] (f3),
(i1) -- [fermion] (a),
(a)-- [red,boson, edge label'=\(\gamma\)] (c)-- [fermion] (f2),
(c) -- [fermion] (f1), 
(a)-- [red,boson, edge label'=\(\gamma\)] (b)
};

\end{feynman}

\end{tikzpicture}
}

\caption{Diagrams (of type II on the left and type III on the right) contributing to the amplitude \eqref{a+a+->b-b-b+b+} which separately diverge in the massless limit $\omega\rightarrow 0$. However, the divergent parts cancel when the contributions of  both diagrams are summed. Their contribution is evaluated in \eqref{A1} and \eqref{A2}, respectively.}
 \label{A12}
\end{figure}
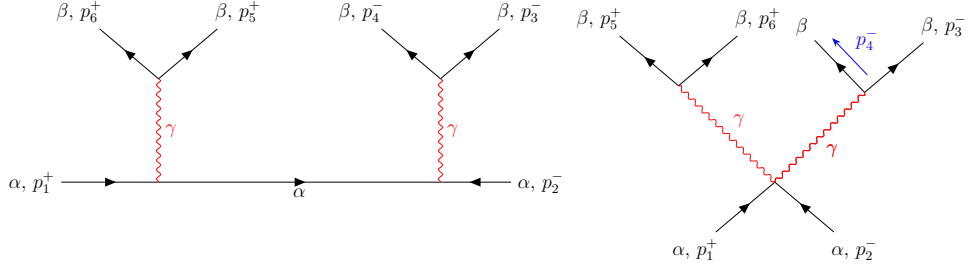
\subsubsection{Diagrams of type II and III }
The diagrams of type II are constituted by four 3-point vertices appearing in \eqref{Lag1}, while the diagrams of type III are constituted by two 3-point vertices appearing in \eqref{Lag1} and one vertex coming from the second term in \eqref{Lag2}. These are depicted on the right and left of fig. \ref{A12}, respectively.
Proceeding, the left diagram in fig. \ref{A12} gives the following result
 \begin{eqnarray}\label{A1}
A_1&=&\omega^4 2^4 \l^2 \tilde D(p_3+p_4)\tilde D(p_5+p_6) D(p_1-p_5-p_6)\,(p_5^0+p_6^0)^2(p_3^0+p_4^0)^2\overset{\omega\rightarrow 0}{=}\nonumber \\ 
&& 4 i \l^2\left(\frac{4 p_3 p_4 p_5 p_6}{\omega^2}-\frac{\left(p_3^2+p_3 p_4+p_4^2\right) \left(p_5^2+p_5 p_6+p_6^2\right)}{(p_3+p_4) (p_5+p_6)}+\frac{p_4 p_5 p_6}{p_3}\right. + \nonumber \\ 
&&\left. p_3 \left(\frac{p_4 (p_5+p_6)^2}{p_5 p_6}+\frac{p_5 p_6}{p_4}\right)+2 p_5 p_6\right).
\end{eqnarray}
The right diagram in fig. \ref{A12} gives
\begin{eqnarray}\label{A2}
&&A_2=\omega^2 2^3 i \l^2\tilde D(p_3+p_4)\tilde D(p_5+p_6)\,(p_5+p_6)\cdot (p_3+p_4) \,(p_5^0+p_6^0)(p_3^0+p_4^0)
\overset{\omega\rightarrow 0}{=} \nonumber \\
&&-16 i \l^2\frac{\omega ^2 p_6 p_5 p_3 p_4}{m^4}-\frac{\omega ^2 \l^24 i \left(\frac{p_6 p_4 p_5}{p_3}+p_3 \left(\frac{p_4 p_5}{p_6}+\frac{p_4 p_6}{p_5}+\frac{p_5 p_6}{p_4}+p_4\right)+p_5 p_6\right)}{m^2}
=\nonumber \\
&&-16 i \l^2\frac{ p_6 p_5 p_3 p_4}{\omega^2}-4 i \l^2\left(\frac{p_6 p_4 p_5}{p_3}+p_3 \left(\frac{p_4 p_5}{p_6}+\frac{p_4 p_6}{p_5}+\frac{p_5 p_6}{p_4}+p_4\right)+p_5 p_6\right)
\end{eqnarray}

At this point let us stress that although both $A_1$ and  $A_2$ diverge in the limit $\omega\rightarrow 0$ their sum is finite, namely
\begin{eqnarray}\label{Am12n}
&&A_1+A_2=4 i \l^2\left(-\frac{p_5^2 \left(p_3^2+p_3 p_4+p_4^2\right)}{(p_3+p_4) (p_5+p_6)}-\frac{p_6 \left(p_3^2-p_5 (p_3+p_4)+p_3 p_4+p_4^2\right)}{p_3+p_4}+p_3 p_4\right)\nonumber \\
&&=4 i \l^2\left(-\frac{p_5^2 \left(p_3^2+p_3 p_4+p_4^2\right)}{p_2 p_1}-\frac{p_6 \left(p_3^2-p_5 p_2+p_3 p_4+p_4^2\right)}{p_2}+p_3 p_4\right)
\end{eqnarray}
Finally, let us mention that the momenta appearing in all the expressions after taking the $\omega\rightarrow 0$ limit are the momenta along the one-dimensional space direction.
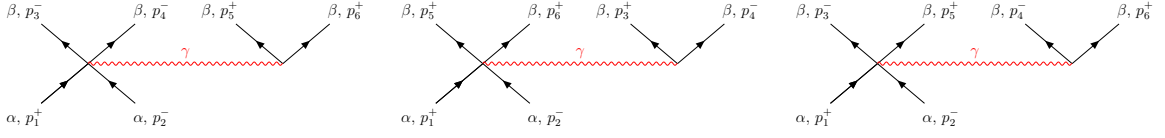
\begin{figure}

\resizebox{5cm}{!}{
\begin{tikzpicture}

\begin{feynman}
\vertex (a);
\vertex [left=5.2cm of a](b) ;
\vertex [above right=of a](f1) {\(\b\), \( p_6^+\)};
\vertex [above left=of a](f2) {\(\b\), \( p_5^+\)};
\vertex [above right=of b](f3) {\(\b\), \( p_4^-\)};
\vertex [above left=of b](f4) {\(\b\), \( p_3^-\)};
\vertex [below right=of b](i2) {\(\a\), \( p_2^-\)};
\vertex [below left=of b](i1) {\(\a\), \( p_1^+\)};

 \diagram* {
(i1) -- [fermion] (b)-- [fermion] (f3),
(b)-- [fermion] (f4),
(i1) -- [fermion] (b),
(i2) -- [fermion] (b),
(a) -- [fermion] (f1),
(a) -- [fermion] (f2),
(a) --[red,boson, edge label'=\(\gamma\)] (b)
}; 

\hspace{300pt}

\vertex (a);
\vertex [left=5.2cm of a](b) ;
\vertex [above right=of a](f1) {\(\b\), \( p_4^-\)};
\vertex [above left=of a](f2) {\(\b\), \( p_3^+\)};
\vertex [above right=of b](f3) {\(\b\), \( p_6^+\)};
\vertex [above left=of b](f4) {\(\b\), \( p_5^+\)};
\vertex [below right=of b](i2) {\(\a\), \( p_2^-\)};
\vertex [below left=of b](i1) {\(\a\), \( p_1^+\)};

 \diagram* {
(i1) -- [fermion] (b)-- [fermion] (f3),
(b)-- [fermion] (f4),
(i1) -- [fermion] (b),
(i2) -- [fermion] (b),
(a) -- [fermion] (f1),
(a) -- [fermion] (f2),
(a) --[red,boson, edge label'=\(\gamma\)] (b)
}; 

 
\hspace{300pt}

\vertex (a);
\vertex [left=5.2cm of a](b) ;
\vertex [above right=of a](f1) {\(\b\), \( p_6^+\)};
\vertex [above left=of a](f2) {\(\b\), \( p_4^-\)};
\vertex [above right=of b](f3) {\(\b\), \( p_5^+\)};
\vertex [above left=of b](f4) {\(\b\), \( p_3^-\)};
\vertex [below right=of b](i2) {\(\a\), \( p_2^-\)};
\vertex [below left=of b](i1) {\(\a\), \( p_1^+\)};

 \diagram* {
(i1) -- [fermion] (b)-- [fermion] (f3),
(b)-- [fermion] (f4),
(i1) -- [fermion] (b),
(i2) -- [fermion] (b),
(a) -- [fermion] (f1),
(a) -- [fermion] (f2),
(a) --[red,boson, edge label'=\(\gamma\)] (b)
};

\end{feynman}

\end{tikzpicture}
}

\caption{Diagrams of type IV involving a 5-point and a 3-point vertex, each proportional to $\omega$. By counting powers of  $\omega$ one can see that the rightmost diagram is 0 in the limit 
$\omega\rightarrow 0$. The same is true for the 3 additional diagrams, similar to the rightmost one, which are obtained by cyclic  permutations of the $\b$ external particles. The contribution of the left and middle diagrams is evaluated in \eqref{A3}.}
 \label{A3f}
\end{figure}

\subsubsection{Diagrams of type IV }
The diagrams of type IV are those involving one vertex appearing in \eqref{Lag1} and one vertex appearing in \eqref{Lag3}. These are depicted in fig. \ref{A3f}.
Next we calculate their contribution. It reads
\begin{eqnarray}\label{A3}
A_3&=&2^3 \omega^2 \l^2\Big( \tilde D(p_5+p_6)\,(p_5^0+p_6^0)^2+\tilde D(p_3+p_4)\,(p_3^0+p_4^0)^2\Big)\overset{\omega\rightarrow 0}{=}\nonumber \\
&&
8 i\l^2\,( p_5 p_6+p_3 p_4)
\end{eqnarray}
\subsubsection{More diagrams of type III }
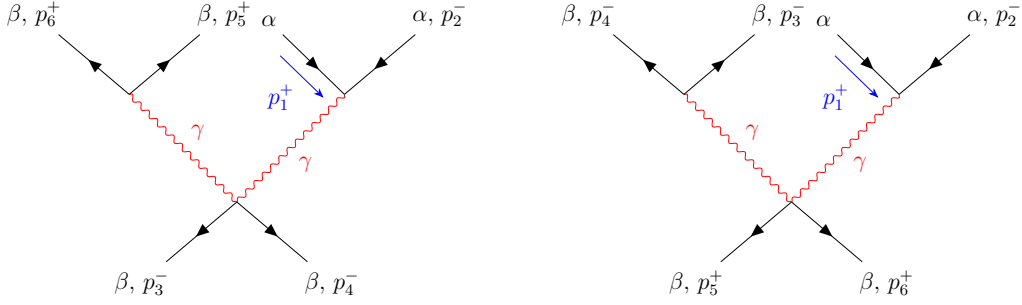
\begin{figure}

\resizebox{6.5cm}{!}{
\begin{tikzpicture}

\begin{feynman}
\vertex (a);
\vertex [below left=of a](i1) {\(\b\), \( p_3^-\)};
\vertex [below right=of a](i2) {\(\b\), \( p_4^-\)};
\vertex [above right=2.7cm of a](b);
\vertex [above left=2.7cm of a](c);
\vertex [above left=of b](f3) {\( \a\)}; 
\vertex [above right=of b](f4) {\(\a\), \(p_2^-\)};
\vertex [above right=of c](f2) {\(\beta\), \( p_5^+\)}; 
\vertex [above left=of c](f1) {\(\beta\), \( p_6^+\)};

 \diagram* {
 (a)--[fermion] (i2),
 (a)--[fermion] (i1),
 (a)-- [red,boson, edge label'=\(\gamma\)] (b),
 (f4)-- [fermion] (b),
(f3)-- [fermion, momentum'={[arrow style=blue]\(p_1^+\)}] (b),
(a)-- [red,boson, edge label'=\(\gamma\)] (c)-- [fermion] (f2),
(c) -- [fermion] (f1), 
}; 

\hspace{280pt}

\vertex (a);
\vertex [below left=of a](i1) {\(\b\), \( p_5^+\)};
\vertex [below right=of a](i2) {\(\b\), \( p_6^+\)};
\vertex [above right=2.7cm of a](b);
\vertex [above left=2.7cm of a](c);
\vertex [above left=of b](f3) {\( \a\)}; 
\vertex [above right=of b](f4) {\(\a\), \(p_2^-\)};
\vertex [above right=of c](f2) {\(\beta\), \( p_3^-\)}; 
\vertex [above left=of c](f1) {\(\beta\), \( p_4^-\)};

 \diagram* {
 (a)--[fermion] (i2),
 (a)--[fermion] (i1),
 (a)-- [red,boson, edge label'=\(\gamma\)] (b),
 (f4)-- [fermion] (b),
(f3)-- [fermion, momentum'={[arrow style=blue]\(p_1^+\)}] (b),
(a)-- [red,boson, edge label'=\(\gamma\)] (c)-- [fermion] (f2),
(c) -- [fermion] (f1), 
}; 
\end{feynman}

\end{tikzpicture}
}

\caption{Seagull diagrams involving 2 vertices of the type \eqref{Lag1} 
and one vertex of the type $-\frac{1}{2}(\alpha^2+\beta^2)\partial_{\mu}\gamma   \partial^{\mu}\gamma$ appearing in \eqref{Lag2}.}
 \label{A4f}
\end{figure}
We now evaluate the contribution of the diagrams in fig. \ref{A4f}. It reads
\begin{eqnarray}
&&A_4=\omega^2 2^3 i \l^2\tilde D(p_5+p_6)\tilde D(p_1+p_2)\,(p_5+p_6)\cdot (p_1+p_2) \,(p_5^0+p_6^0)(p_1^0+p_2^0)+\nonumber \\
&&\omega^2 2^3 i \l^2 \tilde D(p_3+p_4)\tilde D(p_1+p_2)\,(p_3+p_4)\cdot (p_1+p_2) \,(p_3^0+p_4^0)(p_1^0+p_2^0)
\overset{\omega\rightarrow 0}{=}
\nonumber \\
&&
-4 i\l^2 (p_1+p_2) \left(\frac{ p_5 p_6}{p_1}+\frac{ p_3 p_4}{p_2}\right)
\end{eqnarray}
\subsubsection{Diagrams of type V }
\begin{figure}

\resizebox{7.2cm}{!}{
\begin{tikzpicture}

\begin{feynman}
\vertex (a);
\vertex [below=of a](i1) {\(\a\), \( p_1^+\)};
\vertex [left=of a](i2) {\(\a\), \( p_2^-\)};
\vertex [above=2.7cm of a](f4){\(\beta\), \( p_6^+\)};
\vertex [right=3.5cm of a](b);
\vertex [right=of b](f1){\(\beta\), \( p_5^+\)}; 
\vertex [above=of b](c);
\vertex [above right=of c](f2){\(\beta\), \( p_3^-\)};
\vertex [above left=of c](f3){\(\beta\), \( p_4^-\)};

 \diagram* {
 (i1)--[fermion] (a)--[fermion, edge label'=\(\b\)](b)--[fermion](f1),
 (i2)--[fermion] (a)--[fermion] (f4),
 (b)-- [red,boson, edge label'=\(\gamma\)] (c)--[fermion](f2),
 (c)--[fermion](f3),
}; 

\hspace{280pt}

\vertex (a);
\vertex [below=of a](i1) {\(\a\), \( p_1^+\)};
\vertex [left=of a](i2) {\(\a\), \( p_2^-\)};
\vertex [above=2.7cm of a](f4){\(\beta\), \( p_4^-\)};
\vertex [right=3.5cm of a](b);
\vertex [right=of b](f1){\(\beta\), \( p_3^-\)}; 
\vertex [above=of b](c);
\vertex [above right=of c](f2){\(\beta\), \( p_6^+\)};
\vertex [above left=of c](f3){\(\beta\), \( p_5^+\)};

 \diagram* {
 (i1)--[fermion] (a)--[fermion, edge label'=\(\b\)](b)--[fermion](f1),
 (i2)--[fermion] (a)--[fermion] (f4),
 (b)-- [red,boson, edge label'=\(\gamma\)] (c)--[fermion](f2),
 (c)--[fermion](f3),
}; 
\end{feynman}

\end{tikzpicture}
}

\caption{Diagrams of type V with a non-zero contribution in the limit $\omega\rightarrow 0$. One should add 2 additional diagrams of this type. One like the left diagram with $p_5\leftrightarrow p_6$ and one like the right diagram  with $p_3\leftrightarrow p_4$. Their contribution is calculated in \eqref{A5}. The left vertex can only be $-{1 \over 2}\a^2\partial_\mu\b \partial^\mu\b$. In the case  in which the left vertex is ${\omega^2 \over 2}\a^2\b^2$ the corresponding diagrams go to 0 since they are proportional to $\omega^2$. }
 \label{A5f}
\end{figure}
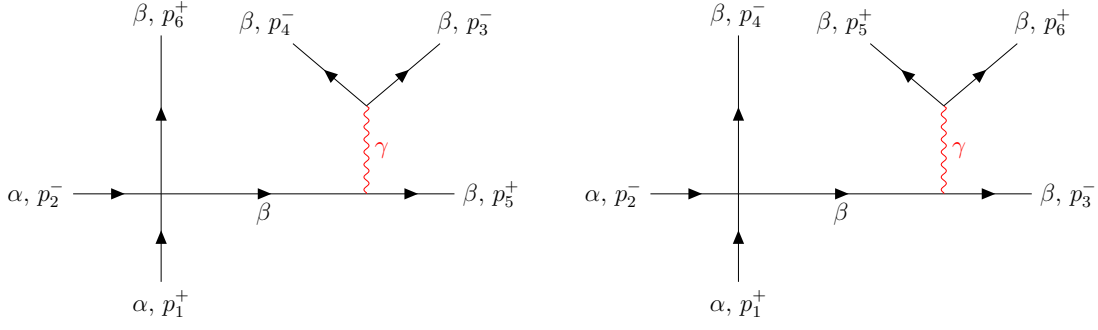

The diagrams of type V involve two vertices from \eqref{Lag1} and one vertex which can be either 
${\omega^2 \over 2}\a^2\b^2$ or $-{1 \over 2}\a^2\partial_\mu\b \partial^\mu\b$.
In what follows, we calculate the contribution of the diagrams in fig. \ref{A5f}.
\begin{eqnarray}\label{A5}
&&A_5=-\omega^2 2^3 i \l^2\tilde D(p_3+p_4)D(p_1+p_2-p_6)\,p_6\cdot (p_1+p_2-p_6) \,(p_3^0+p_4^0)^2+(p_5\leftrightarrow p_6)\nonumber \\
&&-\omega^2 2^3 i \l^2\tilde D(p_5+p_6)D(p_1+p_2-p_4)\,p_4\cdot (p_1+p_2-p_4) \,(p_5^0+p_6^0)^2+(p_3\leftrightarrow p_4)\overset{\omega\rightarrow 0}{=}\nonumber \\
&&
4 i\l^2\Bigg(\frac{  p_5 p_6 \left(p_3^2+p_4^2\right)}{p_3 p_4 }+\frac{  p_3 p_4 \left(p_5^2+p_6^2\right)}{p_5 p_6 }\Bigg)
\end{eqnarray}

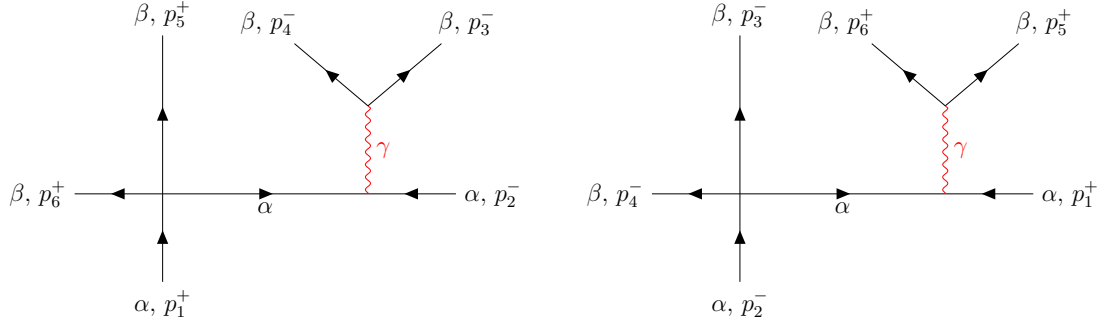
\begin{figure}
\resizebox{7.2cm}{!}{
\begin{tikzpicture}

\begin{feynman}
\vertex (a);
\vertex [below=of a](i1) {\(\a\), \( p_1^+\)};
\vertex [left=of a](i2) {\(\b\), \( p_6^+\)};
\vertex [above=2.7cm of a](f4){\(\beta\), \( p_5^+\)};
\vertex [right=3.5cm of a](b);
\vertex [right=of b](f1){\(\a\), \( p_2^-\)}; 
\vertex [above=of b](c);
\vertex [above right=of c](f2){\(\beta\), \( p_3^-\)};
\vertex [above left=of c](f3){\(\beta\), \( p_4^-\)};

 \diagram* {
 (i1)--[fermion] (a)--[fermion, edge label'=\(\a\)](b),
 (f1)--[fermion](b),
 (a)--[fermion] (f4),
 (a)--[fermion] (i2),
 (b)-- [red,boson, edge label'=\(\gamma\)] (c)--[fermion](f2),
 (c)--[fermion](f3),
}; 

\hspace{280pt}

\vertex (a);
\vertex [below=of a](i1) {\(\a\), \( p_2^-\)};
\vertex [left=of a](i2) {\(\b\), \( p_4^-\)};
\vertex [above=2.7cm of a](f4){\(\beta\), \( p_3^-\)};
\vertex [right=3.5cm of a](b);
\vertex [right=of b](f1){\(\a\), \( p_1^+\)}; 
\vertex [above=of b](c);
\vertex [above right=of c](f2){\(\beta\), \( p_5^+\)};
\vertex [above left=of c](f3){\(\beta\), \( p_6^+\)};

 \diagram* {
 (i1)--[fermion] (a)--[fermion, edge label'=\(\a\)](b),
 (f1)--[fermion](b),
 (a)--[fermion] (f4),
 (a)--[fermion] (i2),
 (b)-- [red,boson, edge label'=\(\gamma\)] (c)--[fermion](f2),
 (c)--[fermion](f3),
}; 
\end{feynman}

\end{tikzpicture}
}

\caption{Diagrams of type V with a non-zero contribution in the limit $\omega\rightarrow 0$. The left vertex can be either ${\omega^2 \over 2}\a^2\b^2$ or $-{1 \over 2}\a^2\partial_\mu\b \partial^\mu\b$ giving the contributions $A_6$ and $A_7$, respectively.}
\label{A67f}
\end{figure}
The contribution of the diagrams shown in fig. \ref{A67f} is given by
\begin{eqnarray}
&&A_6=-\omega^4 2^3 i \l^2\tilde D(p_3+p_4)D(p_1-p_5-p_6) \,(p_3^0+p_4^0)^2\nonumber \\
&&-\omega^4 2^3 i \l^2\tilde D(p_5+p_6)D(p_2-p_3-p_4) \,(p_5^0+p_6^0)^2\overset{\omega\rightarrow 0}{=}\nonumber \\
&&-8 i \l^2\left(\frac{p_1^2 p_5 p_6}{(p_5+p_6)^2}+\frac{p_2^2 p_3 p_4}{(p_3+p_4)^2}\right)=-8 i\l^2\,( p_5 p_6+p_3 p_4)
\end{eqnarray}
and
\begin{eqnarray}
&&A_7=-\omega^2 2^3 i \l^2\tilde D(p_3+p_4)D(p_1-p_5-p_6) \,p_5\cdot p_6\,(p_3^0+p_4^0)^2\nonumber \\
&&-\omega^2 2^3 i \l^2\tilde D(p_5+p_6)D(p_2-p_3-p_4) \,p_3\cdot p_4\,(p_5^0+p_6^0)^2\overset{\omega\rightarrow 0}{=} \\
&&
-4i\l^2\Bigg( \frac{ p_5 p_6 \left(p_3^2+p_4^2\right)}{p_3 p_4 } +\frac{ p_3 p_4 \left(p_5^2+p_6^2\right)}{p_5 p_6 }\Bigg)\nonumber 
\end{eqnarray}

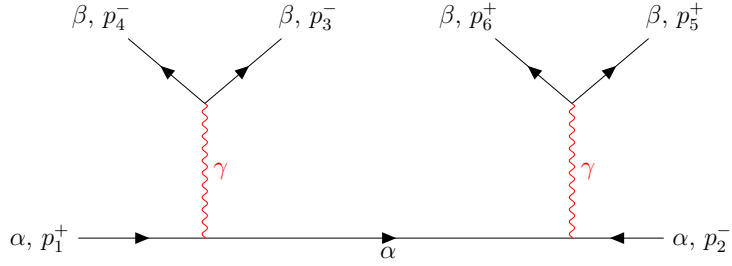
\begin{figure}
\centering

\resizebox{10cm}{!}{
\begin{tikzpicture} 
\begin{feynman}
\vertex (i1){\(\alpha\), \( p_1^+\)};
\vertex [right=2.7 of i1](a);
\vertex [right=6.0cm of a](b);
\vertex [right=of b](i2){\(\alpha\), \( p_2^-\)};
\vertex [above=2.2cm of a](c);
\vertex [above=2.2cm of b](d);
\vertex [above right=of d](f1) {\(\beta\), \(p_5^+\)};
\vertex [above left=of d](f2) {\(\beta\), \(p_6^+\)};
\vertex [above right=of c](f3) {\(\beta\), \(p_3^-\)};
\vertex [above left=of c](f4) {\(\beta\), \(p_4^-\)};

\diagram* {
(i1) -- [fermion] (a)--  [fermion, edge label'=\(\a\)] (b),
(i2) -- [fermion] (b),
(a)-- [red,boson, edge label'=\(\gamma\)] (c)-- [fermion] (f3),
(c) -- [fermion] (f4),
(b)-- [red,boson, edge label'=\(\gamma\)] (d)-- [fermion] (f1),
(d) -- [fermion] (f2),
}; 

\end{feynman}

\end{tikzpicture}
}

\caption{Diagram with 4 vertices of the type \eqref{Lag1}.}
\label{A8f}
\end{figure}
\subsubsection{A last diagram of type II }
Finally, we evaluate the contribution of a last diagram of type II shown in fig. \ref{A8f}. It reads
\begin{eqnarray}
A_8&=&2^4 \omega^4\l^2 \tilde D(p_3+p_4) \tilde D(p_5+p_6) D(p_1-p_3-p_4)(p_3^0+p_4^0)^2 (p_5^0+p_6^0)^2\overset{\omega\rightarrow 0}{=} \nonumber \\
&&\frac{4 i \l^2p_1^2 p_2^2 p_3 p_4 p_5 p_6 \omega ^4}{m^4 (p_3+p_4)^3 (p_5+p_6)^3}
\overset{m=\omega}{=}4i \l^2\frac{ p_3 p_4 p_5 p_6}{p_1 p_2}
\end{eqnarray}

Summing all contributions one gets for the complete amplitude
\be\label{no-part}
A_{\a\a\rightarrow\b\b\b\b} =\sum_{i=0}^8 A_i= 0.
\ee
In conclusion, we have shown that our prescription which was articulated in section \ref{DEF} gives a zero result for the amplitude \eqref{a+a+->b-b-b+b+}. Consequently, the massless S-matrix of the $SU(2)$ PCM does not exhibit particle production as it should be for an integrable theory.
\subsection{A $2\rightarrow 3$ amplitude}
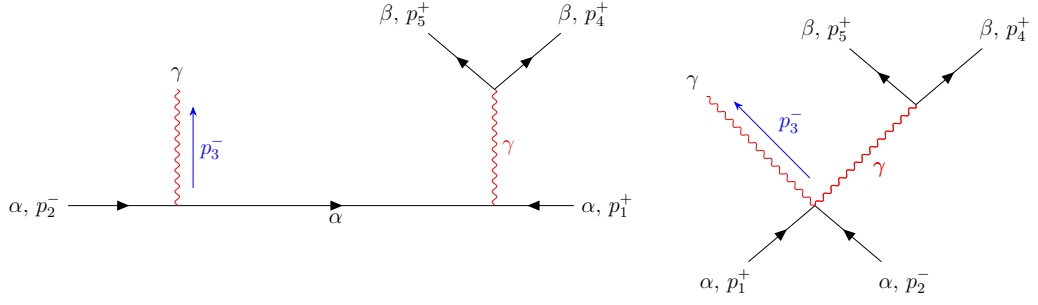
\begin{figure}

\resizebox{10cm}{!}{
\begin{tikzpicture} 
\begin{feynman}
\vertex (i1){\(\alpha\), \( p_2^-\)};
\vertex [right=2.7 of i1](a);
\vertex [right=6.0cm of a](b);
\vertex [right=of b](i2){\(\alpha\), \( p_1^+\)};
\vertex [above=2.2cm of a](c){\(\g\)};
\vertex [above=2.2cm of b](d);
\vertex [above right=of d](f1) {\(\beta\), \(p_4^+\)};
\vertex [above left=of d](f2) {\(\beta\), \(p_5^+\)};

\diagram* {
(i1) -- [fermion] (a)--  [fermion, edge label'=\(\a\)] (b),
(i2) -- [fermion] (b),
(a)-- [red,boson, momentum'={[arrow style=blue]\(p_3^-\)}] (c),
(b)-- [red,boson, edge label'=\(\gamma\)] (d)-- [fermion] (f1),
(d) -- [fermion] (f2),
}; 

\hspace{420pt}. 
\vertex (a);
\vertex [below left=of a](i1) {\(\alpha\), \( p_1^+\)};
\vertex [below right=of a](i2) {\(\alpha\), \( p_2^-\)};
\vertex [above right=2.7cm of a](b);
\vertex [above left=2.9cm of a](c) {\(\g\)};
\vertex [above left=of b](f3) {\( \beta\), \(p_5^+\)}; 
\vertex [above right=of b](f4) {\(\beta\), \(p_4^+\)};

 \diagram* {
(i2) -- [fermion] (a)-- [red,boson, edge label'=\(\gamma\)] (b)-- [fermion] (f4),
(b)-- [fermion] (f3),
(i1) -- [fermion] (a),
(a)-- [red,boson, momentum'={[arrow style=blue]\(p_3^-\)}] (c),
(a)-- [red,boson, edge label'=\(\gamma\)] (b)
};

\end{feynman}

\end{tikzpicture}
}

\caption{Diagrams contributing to the process \eqref{a+a-->g-b+b+}. The two diagrams separately diverge in the massless limit $\omega\rightarrow 0$. However, the total result is zero when the contributions of  both diagrams are summed.}
 \label{Cf}
\end{figure}

In this section, we consider the process in which two particles of the $\a$ species scatter to produce two particles of the $\b$ species and one massless excitation $\g$. More precisely,
\be\label{a+a-->g-b+b+}
\a(p_1^+)\,  \a(p_2^-)\rightarrow \g(p_3^-)\,\b(p_4^+)\,\b(p_5^+)\, .
\ee
This amplitude is zero when using the Lagrangian \eqref{Lagom0}. We will show that it remains zero when using the Lagrangian \eqref{Lag-exp} expanded around the non-trivial vacuum \eqref{nt-vac}, as it should be for our prescription to work. There are two diagrams which contribute in the limit $\omega\rightarrow 0$. These are shown in fig. \ref{Cf}.
The left diagram of fig. \ref{Cf} results to
\begin{eqnarray}
C_1=2^3 \omega^3\l^{3/2}  \tilde D(p_4+p_5) D(p_2-p_3)\, p_3^0\,(p_4^0+p_5^0)^2\,\overset{\omega\rightarrow 0}{=} 
\l^{3/2} \frac{8 p_2 p_4 p_5}{\omega}+\mathcal O(\omega),
\end{eqnarray}
while the right diagram of fig. \ref{Cf} gives
\begin{eqnarray}
C_2=2^2\, i \,\omega\l^{3/2}  \tilde D(p_4+p_5) \, p_3\cdot (p_4+p_5)(p_4^0+p_5^0)\overset{\omega\rightarrow 0}{=} 
-\l^{3/2} \frac{8 p_3 p_4 p_5}{\omega}+\mathcal O(\omega).
\end{eqnarray}
Momentum conservation at leading order implies that $p_2=p_3$. As a result, the sum of the two contributions gives for the full amplitude 
\be\label{no-part-1}
A_{\a\a\rightarrow\g\b\b} =C_1+C_2= 0.
\ee
Before closing this section, let us describe the diagrams each of which gives a zero contribution in the massless limit $\omega\rightarrow 0$. This can be easily verified by counting the powers of $\omega$ in numerators and denomimnators. The first diagram is like the left digram in fig. \ref{Cf} but with the two external $\a$ particles exchanged. The second diagram is also like the left digram in fig. \ref{Cf} but with the two $\a$ excitations exchanged with the two $\b$ excitations. Finally, the third diagram which is zero  in the limit $\omega\rightarrow 0$ is built from a 4-point vertex involving two $\a$ and two $\b$ excitations (one of the two last terms in \eqref{Lag2}) with one of these four particles emitting a photon $\g$.

In conclusion, in this section we have checked that the second point in the highlighted area of section \ref{DEF} is correct.

\section{Equality of the sets of initial and final momenta and factorisation of the S-matrix}
In this section, we will demonstrate that our prescription for the S-matrix is consistent with the equality of the sets of initial and final momenta, as well as with factorisation of the S-matrix.
\subsection{A first  $3\rightarrow 3$ amplitude: equality of the sets of initial and final momenta}\label{sec:6.1}
To this end,  we focus on the following scattering amplitude
\be\label{a+b+b-->a-a+b+}
\a(p_1^+)\,  \a(p_2^+)\b(p_3^-)\,\rightarrow \a(p_4^-)\,\a(p_5^+)\, \b(p_6^+).
\ee

 Energy-momentum conservation with the particles $\a$ and $\b$ being slightly massive  implies
\begin{eqnarray}\label{con-laws2}
&&p_1+p_2=p_5+p_6+a,\, p_3=p_4+a,\\ 
&&a={1\over 16} m^2 \Big(m^2 \big(\sum_{i=1}^3{1\over p_i^3} - \sum_{i=4}^6{1\over p_i^3}\big) + 
   4 \big(-\sum_{i=1}^3{1\over p_i} + \sum_{i=4}^6{1\over p_i}\big)\Big)+\mathcal O(m^6)\nonumber\, .
\end{eqnarray}

\begin{figure} 
\centering

\resizebox{6cm}{!}{
\begin{tikzpicture}


\begin{feynman}
\vertex (a);
\vertex [below=of a](i2) {\(\alpha\), \( p_2^+\)};
\vertex [above=of a](f2) {\(\a\), \( p_5^+\)};
\vertex [below left=of a](i1) {\(\alpha\), \( p_1^+\)};
\vertex [below right=of a](i3) {\(\b\), \( p_3^-\)};
\vertex [above left=of a](f3) {\(\alpha\), \( p_4^-\)};
\vertex [above right=of a](f1) {\(\b\), \( p_6^+\)};

 \diagram* {
(i1) -- [fermion] (a)-- [fermion] (f1),
(a)-- [fermion] (f2),
(a)-- [fermion] (f3),
(i1) -- [fermion] (a),
(i2) -- [fermion] (a),
(i3) -- [fermion] (a),
}; 

\end{feynman}

\end{tikzpicture}
}
\caption{Diagram contributing to the amplitude \ref{a+b+b-->a-a+b+} through a 6-point vertex (last term in \eqref{Lag4}).}
 \label{M0f}
\end{figure}
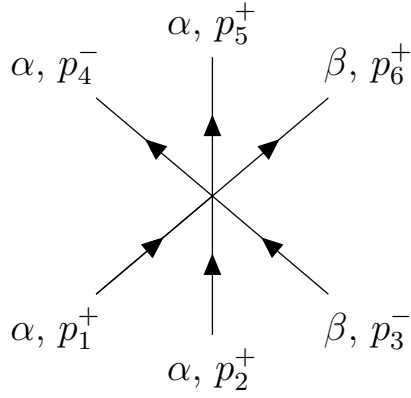
The first contribution to the amplitude \ref{a+b+b-->a-a+b+} is shown in fig. \ref{M0f} and originates from a contact 6-point vertex in \eqref{Lag4}. It is given by
\begin{eqnarray}\label{M0}
&&M_0=8 i \l^2 p_3\cdot p_6\overset{p_3=p_4}{=}8i \l^2 p_4\cdot p_6=16 i \l^2p_4 p_6.
\end{eqnarray}
\subsubsection{Diagrams of type I }
\begin{figure} 
\centering

\resizebox{6cm}{!}{
\begin{tikzpicture}


\begin{feynman}
\vertex (a);
\vertex [below right=of a](i1) {\(\alpha\), \( p_2^+\)};
\vertex [above left =2.7cm of a](f3) {\(\a\), \( p_4^-\)};
\vertex [below left=of a](f4) {\(\b\), \( p_3^-\)};
\vertex [above right=2cm of a](b);
\vertex [above left =of b](f2){\(\a\), \( p_5^+\)};
\vertex [above right=of b](f1){\(\b\), \( p_6^+\)};
\vertex [below right=3cm of b](i2){\(\alpha\), \( p_1^+\)};

 \diagram* {
(i1) -- [fermion] (a)-- [fermion, edge label'=\(\a\)] (b)-- [fermion] (f2),
(b)-- [fermion] (f1),
(i2) -- [fermion] (b),
(a) -- [fermion] (f3),
(f4) -- [fermion] (a),
}; 

\end{feynman}

\end{tikzpicture}
}
\caption{Diagram of type I contributing to the amplitude \ref{a+b+b-->a-a+b+} in the limit $\omega\rightarrow 0$. One should add 5 more diagrams of this type. One  with $p_1\leftrightarrow p_2$,  one with $p_4\leftrightarrow p_5$, one with $p_2\leftrightarrow -p_5$, one with $p_1\leftrightarrow -p_4$ and one with $p_1\leftrightarrow p_2 \land p_4\leftrightarrow p_5$ exchanged.}
 \label{M1f}

\end{figure}
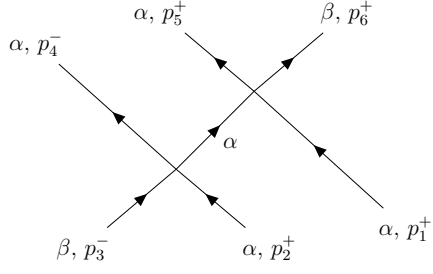
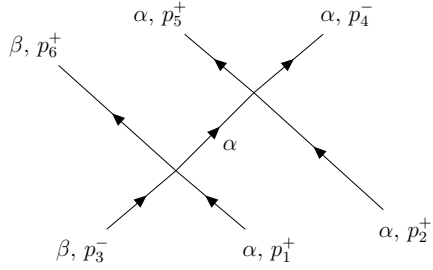
\begin{figure} 
\centering

\resizebox{6cm}{!}{
\begin{tikzpicture}


\begin{feynman}
\vertex (a);
\vertex [below right=of a](i1) {\(\alpha\), \( p_1^+\)};
\vertex [above left =2.7cm of a](f3) {\(\b\), \( p_6^+\)};
\vertex [below left=of a](f4) {\(\b\), \( p_3^-\)};
\vertex [above right=2cm of a](b);
\vertex [above left =of b](f2){\(\a\), \( p_5^+\)};
\vertex [above right=of b](f1){\(\a\), \( p_4^-\)};
\vertex [below right=3cm of b](i2){\(\alpha\), \( p_2^+\)};

 \diagram* {
(i1) -- [fermion] (a)-- [fermion, edge label'=\(\a\)] (b)-- [fermion] (f2),
(b)-- [fermion] (f1),
(i2) -- [fermion] (b),
(a) -- [fermion] (f3),
(f4) -- [fermion] (a),
}; 

\end{feynman}

\end{tikzpicture}
}
\caption{Diagram of type I contributing to the amplitude \ref{a+b+b-->a-a+b+} in the limit $\omega\rightarrow 0$. One should add 3 more diagrams of this type. One  with $p_1\leftrightarrow p_2$, one with $p_1\leftrightarrow -p_4$, one with $p_1\leftrightarrow -p_5$ exchanged.}
 \label{M2f}

\end{figure}

The next contribution originates from the diagrams depicted in fig. \ref{M1f}. Both vertices of the diagram are of the type $-\frac{1}{2}\alpha^2\partial_{\mu}\beta   \partial^{\mu}\beta$. It reads
\begin{eqnarray}\label{M1}
M_1&=&-2^2 \l^2 D(p_2+p_3-p_4)\,\,p_3\cdot (p_2+p_3-p_4)\,  p_6\cdot (p_2+p_3-p_4) +   
\nonumber \\
&&-2^2 \l^2 D(p_1+p_3-p_4)\,\,p_3\cdot (p_1+p_3-p_4)\,  p_6\cdot (p_1+p_3-p_4)
\nonumber \\
&&-2^2 \l^2 D(p_2+p_3-p_5)\,\,p_3\cdot (p_2+p_3-p_5)\,  p_6\cdot (p_2+p_3-p_5)
\nonumber \\
&&-2^2 \l^2 D(p_3-p_4-p_5)\,\,p_3\cdot (p_3-p_4-p_5)\,  p_6\cdot (p_3-p_4-p_5)
\nonumber \\
&&-2^2 \l^2 D(p_1+p_2+p_3)\,\,p_3\cdot (p_1+p_2+p_3)\,  p_6\cdot (p_1+p_2+p_3)
 \\
&&-2^2 \l^2 D(p_1+p_3-p_5)\,\,p_3\cdot (p_1+p_3-p_5)\,  p_6\cdot (p_1+p_3-p_5)
\overset{\omega\rightarrow 0}{=}\nonumber \\
&&4 i \l^2\frac{p_4 p_6 \Big(-\big(p_2^2 (7 p_5+5 p_6)\big)+p_2 (p_5+p_6) (7 p_5+5 p_6)-5 p_5 p_6 (p_5+p_6)\Big)}{(p_2-p_5) (p_2-p_6) (p_5+p_6)}.\nonumber
\end{eqnarray}
Next we write down the result for the diagrams depicted in fig. \ref{M2f}. Here one vertex is of of the type $-\frac{1}{2}\alpha^2\partial_{\mu}\beta   \partial^{\mu}\beta$ and one of the type $\frac{\omega^2}{6}(\alpha^4+\beta^4)$ appearing in \eqref{Lag2}. As mentioned in the corresponding caption there are 3 additional diagrams of this type. 
As a result, one gets
\begin{eqnarray}\label{M2}
M_2&=&8 \omega^2 \l^2 D(p_1+p_3-p_6)\,\,p_3\cdot p_6\, +   (p_1\leftrightarrow p_2)+(p_1\leftrightarrow -p_4)+(p_1\leftrightarrow -p_5)\overset{\omega\rightarrow 0}{=}
\nonumber \\
&&16 i \l^2\frac{ p_4 p_6}{p_6 \left(\frac{1}{p_1}+\frac{1}{p_2}-\frac{1}{p_5}\right)-1}.
\end{eqnarray}

The next contribution originates from the diagrams depicted in fig. \ref{M1f} but with one of the vertices being ${\omega^2 \over 2}\a^2 \b^2$ instead of $-\frac{1}{2}\alpha^2\partial_{\mu}\beta   \partial^{\mu}\beta$. It reads
\begin{eqnarray}\label{M3}
M_3&=&4 \omega^2 \l^2 D(p_2+p_3-p_4)\,\,\Big(p_3\cdot (p_2+p_3-p_4) + p_6\cdot (p_2+p_3-p_4)\Big) +   
\nonumber \\
&&(p_1\leftrightarrow p_2)+(p_4\leftrightarrow p_5)+(p_2\leftrightarrow -p_5)+(p_1\leftrightarrow -p_4)+ (p_1\leftrightarrow p_2 \land p_4\leftrightarrow p_5)\overset{\omega\rightarrow 0}{=}\nonumber \\
&&-24 i \l^2\frac{ p_2 p_4 p_5 p_6 p_1}{(p_2-p_5) (p_2-p_6) (p_5+p_6)}.
\end{eqnarray}
\subsubsection{Diagrams of type II }
\begin{figure}

\resizebox{8cm}{!}{
\begin{tikzpicture} 
\begin{feynman}
\vertex (i1){\(\beta\), \( p_3^-\)};
\vertex [right=2.7 of i1](a);
\vertex [right=6.0cm of a](b);
\vertex [right=of b](i2){\(\beta\), \( p_6^+\)};
\vertex [above=2.2cm of a](c);
\vertex [above=2.2cm of b](d);
\vertex [above right=of d](f1) {\(\a\), \(p_5^+\)};
\vertex [above left=of d](f2) {\(\a\), \(p_4^-\)};
\vertex [above right=of c](f3) {\(\a\), \(p_2^+\)};
\vertex [above left=of c](f4) {\(\a\), \(p_1^+\)};

\diagram* {
(i1) -- [fermion] (a)--  [fermion, edge label'=\(\b\)] (b),
(b) -- [fermion] (i2),
(a)-- [red,boson, edge label'=\(\gamma\)] (c),
(f4) -- [fermion] (c),
(f3) -- [fermion] (c),
(b)-- [red,boson, edge label'=\(\gamma\)] (d)-- [fermion] (f1),
(d) -- [fermion] (f2),
}; 

\hspace{370pt}

\vertex (i1){\(\beta\), \( p_3^-\)};
\vertex [right=2.7 of i1](a);
\vertex [right=6.0cm of a](b);
\vertex [right=of b](i2){\(\beta\), \( p_6^+\)};
\vertex [above=2.2cm of a](c);
\vertex [above=2.2cm of b](d);
\vertex [above right=of d](f1) {\(\a\), \(p_1^+\)};
\vertex [above left=of d](f2) {\(\a\), \(p_4^-\)};
\vertex [above right=of c](f3) {\(\a\), \(p_2^+\)};
\vertex [above left=of c](f4) {\(\a\), \(p_5^+\)};

\diagram* {
(i1) -- [fermion] (a)--  [fermion, edge label'=\(\b\)] (b),
(b) -- [fermion] (i2),
(a)-- [red,boson, edge label'=\(\gamma\)] (c),
(c) -- [fermion] (f4),
(f3) -- [fermion] (c),
(b)-- [red,boson, edge label'=\(\gamma\)] (d),
(f1) -- [fermion] (d),
(d) -- [fermion] (f2),
};

\end{feynman}

\end{tikzpicture}

}

\centering
\resizebox{8cm}{!}{
\begin{tikzpicture} 
\begin{feynman}
\vertex (i1){\(\beta\), \( p_3^-\)};
\vertex [right=2.7 of i1](a);
\vertex [right=6.0cm of a](b);
\vertex [right=of b](i2){\(\beta\), \( p_6^+\)};
\vertex [above=2.2cm of a](c);
\vertex [above=2.2cm of b](d);
\vertex [above right=of d](f1) {\(\a\), \(p_2^+\)};
\vertex [above left=of d](f2) {\(\a\), \(p_4^-\)};
\vertex [above right=of c](f3) {\(\a\), \(p_5^+\)};
\vertex [above left=of c](f4) {\(\a\), \(p_1^+\)};

\diagram* {
(i1) -- [fermion] (a)--  [fermion, edge label'=\(\b\)] (b),
(b) -- [fermion] (i2),
(a)-- [red,boson, edge label'=\(\gamma\)] (c),
(f4) -- [fermion] (c),
(c) -- [fermion] (f3),
(b)-- [red,boson, edge label'=\(\gamma\)] (d),
(f1) -- [fermion] (d),
(d) -- [fermion] (f2),
}; 

\end{feynman}
\end{tikzpicture}
}
\caption{Diagrams  of type II contributing to \eqref{a+b+b-->a-a+b+} in the massless limit $\omega\rightarrow 0$. One should add 3 additional diagrams like the ones shown but with the particles of the $\b$ species exchanged, which amounts to $p_3\leftrightarrow -p_6$ in the expression for each diagram.}
 \label{X1f}
\end{figure}
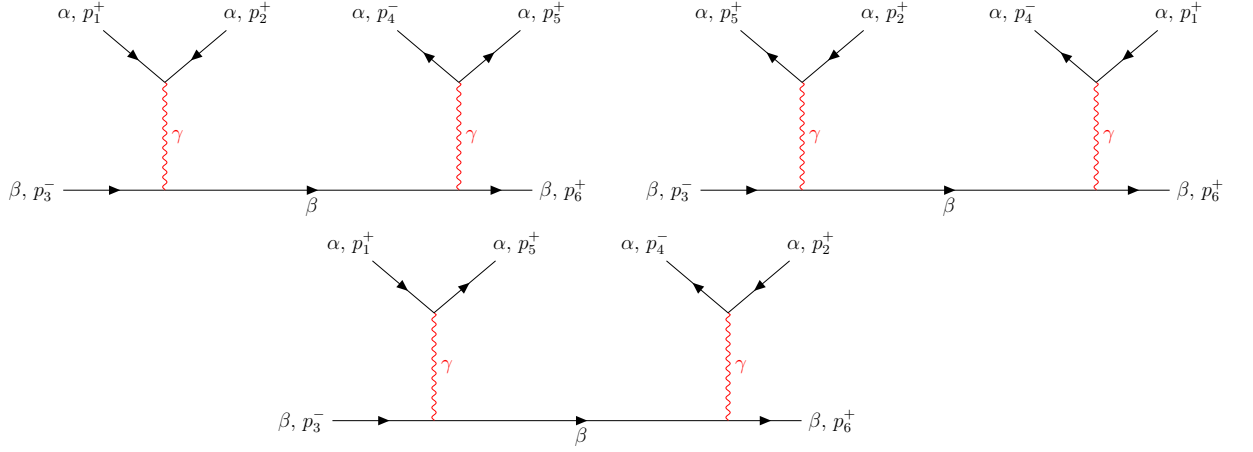

\begin{figure}

\resizebox{8cm}{!}{
\begin{tikzpicture} 
\begin{feynman}
\vertex (i1){\(\a\), \( p_1^+\)};
\vertex [right=2.7 of i1](a);
\vertex [right=6.0cm of a](b);
\vertex [right=of b](i2){\(\a\), \( p_2^+\)};
\vertex [above=2.2cm of a](c);
\vertex [above=2.2cm of b](d);
\vertex [above right=of d](f1) {\(\a\), \(p_5^+\)};
\vertex [above left=of d](f2) {\(\a\), \(p_4^-\)};
\vertex [above right=of c](f3) {\(\b\), \(p_6^+\)};
\vertex [above left=of c](f4) {\(\b\), \(p_3^-\)};

\diagram* {
(i1) -- [fermion] (a)--  [fermion, edge label'=\(\a\)] (b),
(i2) -- [fermion] (b),
(a)-- [red,boson, edge label'=\(\gamma\)] (c),
(f4) -- [fermion] (c),
(c) -- [fermion] (f3),
(b)-- [red,boson, edge label'=\(\gamma\)] (d)-- [fermion] (f1),
(d) -- [fermion] (f2),
}; 

\hspace{370pt}

\vertex (i1){\(\a\), \( p_1^+\)};
\vertex [right=2.7 of i1](a);
\vertex [right=6.0cm of a](b);
\vertex [right=of b](i2){\(\a\), \( p_4^-\)};
\vertex [above=2.2cm of a](c);
\vertex [above=2.2cm of b](d);
\vertex [above right=of d](f1) {\(\a\), \(p_5^+\)};
\vertex [above left=of d](f2) {\(\a\), \(p_2^+\)};
\vertex [above right=of c](f3) {\(\b\), \(p_6^+\)};
\vertex [above left=of c](f4) {\(\b\), \(p_3^-\)};

\diagram* {
(i1) -- [fermion] (a)--  [fermion, edge label'=\(\a\)] (b),
(b) -- [fermion] (i2),
(a)-- [red,boson, edge label'=\(\gamma\)] (c),
(f4) -- [fermion] (c),
(c) -- [fermion] (f3),
(b)-- [red,boson, edge label'=\(\gamma\)] (d)-- [fermion] (f1),
(f2) -- [fermion] (d),
};

\end{feynman}

\end{tikzpicture}

}

\centering
\resizebox{8cm}{!}{
\begin{tikzpicture} 
\begin{feynman}
\vertex (i1){\(\a\), \( p_1^+\)};
\vertex [right=2.7 of i1](a);
\vertex [right=6.0cm of a](b);
\vertex [right=of b](i2){\(\a\), \( p_5^+\)};
\vertex [above=2.2cm of a](c);
\vertex [above=2.2cm of b](d);
\vertex [above right=of d](f1) {\(\a\), \(p_4^-\)};
\vertex [above left=of d](f2) {\(\a\), \(p_2^+\)};
\vertex [above right=of c](f3) {\(\b\), \(p_6^+\)};
\vertex [above left=of c](f4) {\(\b\), \(p_3^-\)};

\diagram* {
(i1) -- [fermion] (a)--  [fermion, edge label'=\(\a\)] (b),
(b) -- [fermion] (i2),
(a)-- [red,boson, edge label'=\(\gamma\)] (c),
(f4) -- [fermion] (c),
(c) -- [fermion] (f3),
(b)-- [red,boson, edge label'=\(\gamma\)] (d)-- [fermion] (f1),
(f2) -- [fermion] (d),
}; 

\end{feynman}
\end{tikzpicture}
}
\caption{Diagrams  of type II contributing to \eqref{a+b+b-->a-a+b+} in the massless limit $\omega\rightarrow 0$. One should add 3 additional diagrams like the ones shown but with the particles of the $\a$ species  of the horizontal line exchanged, which amounts to $p_1\leftrightarrow p_2$ for the top left diagram, 
$p_1\leftrightarrow -p_4$ for the top right diagram and $p_1\leftrightarrow -p_5$ for the bottom one. Their contribution is calculated in \eqref{X21}.}
 \label{X21f}
\end{figure}
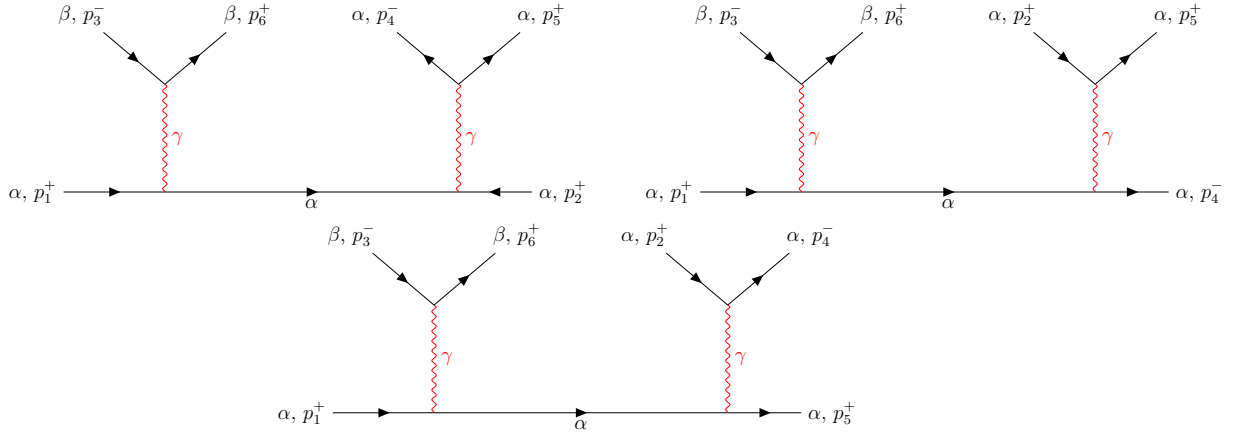

We proceed by calculating the contribution of the diagrams in fig. \ref{X1f}.
\begin{eqnarray}\label{X1}
&&M_4=2^4 \omega^4 \l^2 \Big( D(p_1+p_2+p_3)\tilde D(p_1+p_2)\tilde D(p_4+p_5)\,(p_1^0+p_2^0)^2 \,(p_4^0+p_5^0)^2 +   
\nonumber \\
&&D(p_2+p_3-p_5)\tilde D(p_2-p_5)\tilde D(p_1-p_4)\,(p_2^0-p_5^0)^2 \,(p_1^0-p_4^0)^2+
 \\
&&D(p_1+p_3-p_5)\tilde D(p_1-p_5)\tilde D(p_2-p_4)\,(p_1^0-p_5^0)^2 \,(p_2^0-p_4^0)^2\Big)
\nonumber \\
&&+( \text{3 previous terms with} \,\,\,\,p_3\leftrightarrow -p_6)\overset{\omega\rightarrow 0}{=}
\nonumber \\
&&-\frac{4 i \l^2p_6 \left(\frac{p_1^2 p_2^3 (p_4+p_5)^2 (p_5+p_6)^2 p_1}{(p_1+p_2)^2}-\frac{p_1^2 p_5^3 (p_2-p_4)^2 (p_2-p_6)^2 p_1}{(p_1-p_5)^2}-p_2^3 p_5^3 (p_4-p_1)^2\right)}{p_1 p_2 p_4 p_5 (p_2-p_5) (p_2-p_6) (p_5+p_6)}\nonumber.
\end{eqnarray}
Next we focus on the diagrams of fig. \ref{X21f} and  fig. \ref{X22f}. The former give
\begin{eqnarray}\label{X21}
&&M_5^{(1)}=2^4 \omega^4 \l^2 \Big( D(p_1+p_3-p_6)\tilde D(p_3-p_6)\tilde D(p_4+p_5)\,(p_3^0-p_6^0)^2 \,(p_4^0+p_5^0)^2 +   
\nonumber \\
&&D(p_2+p_3-p_6)\tilde D(p_3-p_6)\tilde D(p_4+p_5)\,(p_3^0-p_6^0)^2 \,(p_4^0+p_5^0)^2 + 
 \\
&&D(p_1+p_3-p_6)\tilde D(p_3-p_6)\tilde D(p_2-p_5)\,(p_3^0-p_6^0)^2 \,(p_2^0-p_5^0)^2 +
\nonumber \\
&&D(p_3-p_4-p_6)\tilde D(p_3-p_6)\tilde D(p_2-p_5)\,(p_3^0-p_6^0)^2 \,(p_2^0-p_5^0)^2 +
\nonumber \\
&& D(p_1+p_3-p_6)\tilde D(p_3-p_6)\tilde D(p_2-p_4)\,(p_3^0-p_6^0)^2 \,(p_2^0-p_4^0)^2 +   
\nonumber \\
&&D(p_3-p_5-p_6)\tilde D(p_3-p_6)\tilde D(p_2-p_4)\,(p_3^0-p_6^0)^2 \,(p_2^0-p_4^0)^2  \Big).\nonumber
\end{eqnarray}
while the latter
\begin{eqnarray}\label{X22}
&&M_5^{(2)}=2^4 \omega^4 \l^2 \Big( D(p_2+p_3-p_6)\tilde D(p_3-p_6)\tilde D(p_1-p_5)\,(p_3^0-p_6^0)^2 \,(p_1^0-p_5^0)^2 +   
\nonumber \\
&&D(p_3-p_4-p_6)\tilde D(p_3-p_6)\tilde D(p_1-p_5)\,(p_3^0-p_6^0)^2 \,(p_1^0-p_5^0)^2 +  
 \\
&&D(p_2+p_3-p_6)\tilde D(p_3-p_6)\tilde D(p_1-p_4)\,(p_3^0-p_6^0)^2 \,(p_1^0-p_4^0)^2 +  
\nonumber \\
&&D(p_3-p_5-p_6)\tilde D(p_3-p_6)\tilde D(p_1-p_4)\,(p_3^0-p_6^0)^2 \,(p_1^0-p_4^0)^2 +
\nonumber \\
&& D(p_3-p_4-p_6)\tilde D(p_3-p_6)\tilde D(p_1+p_2)\,(p_3^0-p_6^0)^2 \,(p_1^0+p_2^0)^2 +   
\nonumber \\
&&D(p_3-p_5-p_6)\tilde D(p_3-p_6)\tilde D(p_1+p_2)\,(p_3^0-p_6^0)^2 \,(p_1^0+p_2^0)^2  \Big).\nonumber
\end{eqnarray}
\begin{figure}

\resizebox{8cm}{!}{
\begin{tikzpicture} 
\begin{feynman}
\vertex (i1){\(\a\), \( p_4^-\)};
\vertex [right=2.7 of i1](a);
\vertex [right=6.0cm of a](b);
\vertex [right=of b](i2){\(\a\), \( p_5^+\)};
\vertex [above=2.2cm of a](c);
\vertex [above=2.2cm of b](d);
\vertex [above right=of d](f1) {\(\a\), \(p_2^+\)};
\vertex [above left=of d](f2) {\(\a\), \(p_1^+\)};
\vertex [above right=of c](f3) {\(\b\), \(p_6^+\)};
\vertex [above left=of c](f4) {\(\b\), \(p_3^-\)};

\diagram* {
[fermion] (a)--  [fermion, edge label'=\(\a\)] (b),
(a) -- [fermion] (i1),
(b) -- [fermion] (i2),
(a)-- [red,boson, edge label'=\(\gamma\)] (c),
(f4) -- [fermion] (c),
(c) -- [fermion] (f3),
(b)-- [red,boson, edge label'=\(\gamma\)] (d),
(f1) -- [fermion] (d),
(f2) -- [fermion] (d),
}; 

\hspace{370pt}

\vertex (i1){\(\a\), \( p_2^+\)};
\vertex [right=2.7 of i1](a);
\vertex [right=6.0cm of a](b);
\vertex [right=of b](i2){\(\a\), \( p_4^-\)};
\vertex [above=2.2cm of a](c);
\vertex [above=2.2cm of b](d);
\vertex [above right=of d](f1) {\(\a\), \(p_5^+\)};
\vertex [above left=of d](f2) {\(\a\), \(p_1^+\)};
\vertex [above right=of c](f3) {\(\b\), \(p_6^+\)};
\vertex [above left=of c](f4) {\(\b\), \(p_3^-\)};

\diagram* {
(i1) -- [fermion] (a)--  [fermion, edge label'=\(\a\)] (b),
(b) -- [fermion] (i2),
(a)-- [red,boson, edge label'=\(\gamma\)] (c),
(f4) -- [fermion] (c),
(c) -- [fermion] (f3),
(b)-- [red,boson, edge label'=\(\gamma\)] (d)-- [fermion] (f1),
(f2) -- [fermion] (d),
};

\end{feynman}

\end{tikzpicture}

}

\centering
\resizebox{8cm}{!}{
\begin{tikzpicture} 
\begin{feynman}
\vertex (i1){\(\a\), \( p_2^+\)};
\vertex [right=2.7 of i1](a);
\vertex [right=6.0cm of a](b);
\vertex [right=of b](i2){\(\a\), \( p_5^+\)};
\vertex [above=2.2cm of a](c);
\vertex [above=2.2cm of b](d);
\vertex [above right=of d](f1) {\(\a\), \(p_4^-\)};
\vertex [above left=of d](f2) {\(\a\), \(p_1^+\)};
\vertex [above right=of c](f3) {\(\b\), \(p_6^+\)};
\vertex [above left=of c](f4) {\(\b\), \(p_3^-\)};

\diagram* {
(i1) -- [fermion] (a)--  [fermion, edge label'=\(\a\)] (b),
(b) -- [fermion] (i2),
(a)-- [red,boson, edge label'=\(\gamma\)] (c),
(f4) -- [fermion] (c),
(c) -- [fermion] (f3),
(b)-- [red,boson, edge label'=\(\gamma\)] (d)-- [fermion] (f1),
(f2) -- [fermion] (d),
}; 

\end{feynman}
\end{tikzpicture}
}
\caption{Diagrams of type II contributing to \eqref{a+b+b-->a-a+b+} in the massless limit $\omega\rightarrow 0$. One should add 3 additional diagrams like the ones shown but with the particles of the $\a$ species  of the horizontal line exchanged, which amounts to $p_4\leftrightarrow p_5$ for the top left diagram, 
$p_2\leftrightarrow -p_4$ for the top right diagram and $p_2\leftrightarrow -p_5$ for the bottom one. Their contribution is calculated in \eqref{X22}.}
 \label{X22f}
\end{figure}

In the massless limit $\omega\rightarrow 0$ the sum of \eqref{X21}and \eqref{X22} results to
\begin{eqnarray}\label{X2}
M_5=M_5^{(1)}+M_5^{(2)}=-4 i \l^2\frac{ p_2 p_5 (p_4-p_6)^2 p_1 \left(p_2^2-p_2 (p_5+p_6)+p_5 (p_5+p_6)\right)}{p_4 p_6 (p_2-p_5) (p_2-p_6) (p_5+p_6)}.\nonumber\\
\end{eqnarray}
\subsubsection{Diagrams of type III}
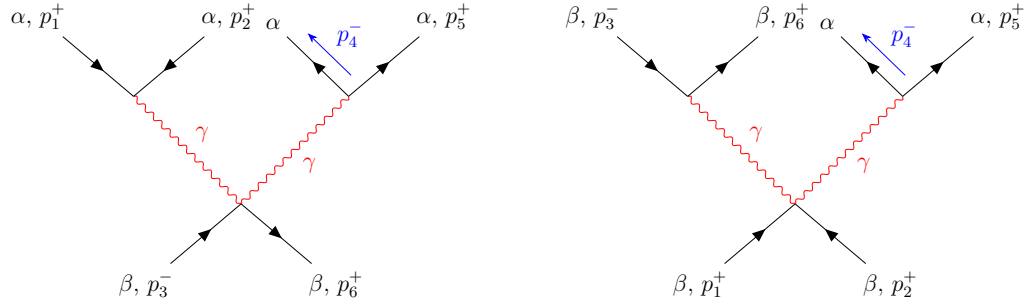
\begin{figure}

\resizebox{6.5cm}{!}{
\begin{tikzpicture}

\begin{feynman}
\vertex (a);
\vertex [below left=of a](i1) {\(\b\), \( p_3^-\)};
\vertex [below right=of a](i2) {\(\b\), \( p_6^+\)};
\vertex [above right=2.7cm of a](b);
\vertex [above left=2.7cm of a](c);
\vertex [above left=of b](f3) {\( \a\)}; 
\vertex [above right=of b](f4) {\(\a\), \(p_5^+\)};
\vertex [above right=of c](f2) {\(\a\), \( p_2^+\)}; 
\vertex [above left=of c](f1) {\(\a\), \( p_1^+\)};

 \diagram* {
 (a)--[fermion] (i2),
 (i1)--[fermion] (a),
 (a)-- [red,boson, edge label'=\(\gamma\)] (b),
 (b)-- [fermion] (f4),
(b)-- [fermion, momentum'={[arrow style=blue]\(p_4^-\)}] (f3),
(a)-- [red,boson, edge label'=\(\gamma\)] (c),
(f2) -- [fermion] (c), 
(f1) -- [fermion] (c), 
}; 

\hspace{280pt}

\vertex (a);
\vertex [below left=of a](i1) {\(\b\), \( p_1^+\)};
\vertex [below right=of a](i2) {\(\b\), \( p_2^+\)};
\vertex [above right=2.7cm of a](b);
\vertex [above left=2.7cm of a](c);
\vertex [above left=of b](f3) {\( \a\)}; 
\vertex [above right=of b](f4) {\(\a\), \(p_5^+\)};
\vertex [above right=of c](f2) {\(\beta\), \( p_6^+\)}; 
\vertex [above left=of c](f1) {\(\beta\), \( p_3^-\)};

 \diagram* {
 (i2)--[fermion] (a),
 (i1)--[fermion] (a),
 (a)-- [red,boson, edge label'=\(\gamma\)] (b),
 (b)-- [fermion] (f4),
(b)-- [fermion, momentum'={[arrow style=blue]\(p_4^-\)}] (f3),
(a)-- [red,boson, edge label'=\(\gamma\)] (c)-- [fermion] (f2),
(f1) -- [fermion] (c), 
}; 
\end{feynman}

\end{tikzpicture}
}

\caption{Seagull diagrams contributing to \eqref{a+b+b-->a-a+b+}  and involving 2 vertices of the type \eqref{Lag1} 
and one vertex of the type $-\frac{1}{2}(\alpha^2+\beta^2)\partial_{\mu}\gamma   \partial^{\mu}\gamma$ appearing in \eqref{Lag2}. One should include 2 more diagrams like the left one but with $p_5\leftrightarrow -p_1$ and $p_5\leftrightarrow -p_2$ exchanged. Moreover, there are 5 additional diagrams similar to the one on the right, one with  $p_2 \leftrightarrow -p_4$, one with  $p_2 \leftrightarrow -p_5$, one with $p_1 \leftrightarrow -p_4$, one with $p_1 \leftrightarrow -p_5$ and one with $p_1 \leftrightarrow -p_4 \land p_2 \leftrightarrow -p_5$ exchanged. The contribution of the left and right diagrams is evaluated in \eqref{Y1} and \eqref{Y2}, respectively.}
 \label{Y12f}
\end{figure}
We now proceed to the diagrams  of fig. \ref{Y12f}. The left diagram with its two companions give
\begin{eqnarray}\label{Y1}
&&M_6=8 \omega^2 \l^2 i \tilde D(p_1+p_2)\tilde D(p_4+p_5)\,(p_1^0+p_2^0)(p_4^0+p_5^0)\,(p_1+p_2)\cdot (p_4+p_5)\, \nonumber\\
&&+  (p_1\leftrightarrow -p_5)+(p_2\leftrightarrow -p_5)\overset{\omega\rightarrow 0}{=}
\nonumber \\
&&4 i \l^2\left(\frac{p_2^2 (p_4+p_5)}{p_5}-p_2 \left(\frac{p_4 p_5}{p_1}+\frac{p_4 p_6}{p_5}+p_4+p_5+p_6\right) \right.
\nonumber \\
&&\left.
-\frac{p_4 p_5 (p_5+p_6)}{p_2}+p_5 (p_4+p_5+p_6)\right),
\end{eqnarray}
while the right diagram of fig. \ref{Y12f} with its five  companions give
\begin{eqnarray}\label{Y2}
&&M_7=8 \omega^2 \l^2 i \tilde D(p_3-p_6)\tilde D(p_4+p_5)\,(p_3^0-p_6^0)(p_4^0+p_5^0)\,(p_3-p_6)\cdot (p_4+p_5)\, +\nonumber\\
&& (p_2\leftrightarrow -p_4)+(p_2\leftrightarrow -p_5)+(p_1\leftrightarrow -p_4)+(p_1\leftrightarrow -p_5)+(p_1\leftrightarrow -p_4\land p_2\leftrightarrow -p_5)\overset{\omega\rightarrow 0}{=}
\nonumber \\
&&-4i \l^2\frac{ (p_4-p_6) \left(p_2^2-p_2 (p_5+p_6)+p_5 (p_5+p_6)\right)}{p_6}.
\end{eqnarray}
\subsubsection{Diagrams of type IV }
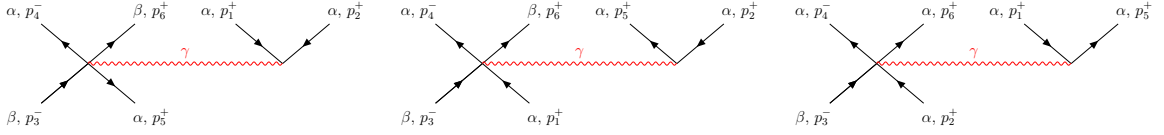
\begin{figure}

\resizebox{5cm}{!}{
\begin{tikzpicture}

\begin{feynman}
\vertex (a);
\vertex [left=5.2cm of a](b) ;
\vertex [above right=of a](f1) {\(\a\), \( p_2^+\)};
\vertex [above left=of a](f2) {\(\a\), \( p_1^+\)};
\vertex [above right=of b](f3) {\(\b\), \( p_6^+\)};
\vertex [above left=of b](f4) {\(\a\), \( p_4^-\)};
\vertex [below right=of b](i2) {\(\a\), \( p_5^+\)};
\vertex [below left=of b](i1) {\(\b\), \( p_3^-\)};

 \diagram* {
(i1) -- [fermion] (b)-- [fermion] (f3),
(b)-- [fermion] (f4),
(i1) -- [fermion] (b),
(b) -- [fermion] (i2),
(f1) -- [fermion] (a),
(f2) -- [fermion] (a),
(a) --[red,boson, edge label'=\(\gamma\)] (b)
}; 

\hspace{300pt}

\vertex (a);
\vertex [left=5.2cm of a](b) ;
\vertex [above right=of a](f1) {\(\a\), \( p_2^+\)};
\vertex [above left=of a](f2) {\(\a\), \( p_5^+\)};
\vertex [above right=of b](f3) {\(\b\), \( p_6^+\)};
\vertex [above left=of b](f4) {\(\a\), \( p_4^-\)};
\vertex [below right=of b](i2) {\(\a\), \( p_1^+\)};
\vertex [below left=of b](i1) {\(\b\), \( p_3^-\)};

 \diagram* {
(i1) -- [fermion] (b)-- [fermion] (f3),
(b)-- [fermion] (f4),
(i1) -- [fermion] (b),
(i2) -- [fermion] (b),
(f1) -- [fermion] (a),
(a) -- [fermion] (f2),
(a) --[red,boson, edge label'=\(\gamma\)] (b)
}; 

 
\hspace{300pt}

\vertex (a);
\vertex [left=5.2cm of a](b) ;
\vertex [above right=of a](f1) {\(\a\), \( p_5^+\)};
\vertex [above left=of a](f2) {\(\a\), \( p_1^+\)};
\vertex [above right=of b](f3) {\(\a\), \( p_6^+\)};
\vertex [above left=of b](f4) {\(\a\), \( p_4^-\)};
\vertex [below right=of b](i2) {\(\a\), \( p_2^+\)};
\vertex [below left=of b](i1) {\(\b\), \( p_3^-\)};

 \diagram* {
(i1) -- [fermion] (b)-- [fermion] (f3),
(b)-- [fermion] (f4),
(i1) -- [fermion] (b),
(i2) -- [fermion] (b),
(a) -- [fermion] (f1),
(f2) -- [fermion] (a),
(a) --[red,boson, edge label'=\(\gamma\)] (b)
};

\end{feynman}

\end{tikzpicture}
}

\caption{Diagrams of type IV contributing to \eqref{a+b+b-->a-a+b+} and involving a 5-point and a 3-point vertex, each proportional to $\omega$.  Their  contribution is evaluated in \eqref{Z}.}
 \label{Zf}
\end{figure}
The three diagrams of fig. \ref{Zf} give the following contribution
\begin{eqnarray}\label{Z}
&&M_8=8 \omega^2 \l^2 \Big(\tilde D(p_1+p_2)\,(p_1^0+p_2^0)^2+\tilde D(p_2-p_5)\,(p_2^0-p_5^0)^2+\tilde D(p_1-p_5)\,(p_1^0-p_5^0)^2\Big)\nonumber\\
&&\overset{\omega\rightarrow 0}{=}-8 i \l^2\left(p_2^2-p_2 (p_5+p_6)+p_5 (p_5+p_6)\right).
\end{eqnarray}
Before continuing, let us note that there are four additional diagrams of the type shown in fig. \ref{Zf} but one can easily check that their contribution vanish in the limit $\omega\rightarrow 0$ for the kinematics that we have chosen. The first of these non-contributing diagrams is the one with both the $\b$ particles being attached to the right vertex of the diagrams in fig. \ref{Zf}. The second one is like the left diagram of fig. \ref{Zf} with the particles having momenta $p_1\leftrightarrow -p_4$ exchanged, the third one  is like the left diagram of fig. \ref{Zf} with the particles having momenta $p_2\leftrightarrow -p_4$ exchanged, and finally the fourth one is like the left diagram of fig. \ref{Zf} with the particles having momenta $p_1\leftrightarrow -p_5$ and  $p_2\leftrightarrow -p_4$ exchanged at the same time.

\subsubsection{Diagrams of type V }
\begin{figure}

\resizebox{7.2cm}{!}{
\begin{tikzpicture}

\begin{feynman}
\vertex (a);
\vertex [below=of a](i1) {\(\b\), \( p_6^+\)};
\vertex [left=of a](i2) {\(\b\), \( p_3^-\)};
\vertex [above=2.7cm of a](f4){\(\a\), \( p_1^+\)};
\vertex [right=3.5cm of a](b);
\vertex [right=of b](f1){\(\a\), \( p_2^+\)}; 
\vertex [above=of b](c);
\vertex [above right=of c](f2){\(\a\), \( p_5^+\)};
\vertex [above left=of c](f3){\(\a\), \( p_4^-\)};

 \diagram* {
 [fermion] (a)--[fermion, edge label'=\(\a\)](b),
 (f1)--[fermion](b),
 (a)--[fermion](i1),
 (i2)--[fermion] (a),
 (f4)--[fermion](a),
 (b)-- [red,boson, edge label'=\(\gamma\)] (c)--[fermion](f2),
 (c)--[fermion](f3),
}; 

\hspace{280pt}

\vertex (a);
\vertex [below=of a](i1) {\(\b\), \( p_6^+\)};
\vertex [left=of a](i2) {\(\b\), \( p_3^-\)};
\vertex [above=2.7cm of a](f4){\(\a\), \( p_2^+\)};
\vertex [right=3.5cm of a](b);
\vertex [right=of b](f1){\(\a\), \( p_1^+\)}; 
\vertex [above=of b](c);
\vertex [above right=of c](f2){\(\a\), \( p_5^+\)};
\vertex [above left=of c](f3){\(\a\), \( p_4^-\)};

 \diagram* {
 [fermion] (a)--[fermion, edge label'=\(\a\)](b),
 (f1)--[fermion](b),
 (a)--[fermion](i1),
 (i2)--[fermion] (a),
 (f4)--[fermion](a),
 (b)-- [red,boson, edge label'=\(\gamma\)] (c)--[fermion](f2),
 (c)--[fermion](f3),
}; 
\end{feynman}

\end{tikzpicture}
}

\caption{Diagrams of type V with a non-zero contribution in the limit $\omega\rightarrow 0$ for the process  \eqref{a+b+b-->a-a+b+}. One should add 4 additional diagrams of this type. Two like the left diagram, one with $p_2\leftrightarrow -p_4$ and one with $p_2\leftrightarrow -p_5$ 
and two like the right diagram, one  with $p_1\leftrightarrow -p_5$ and one with $p_1\leftrightarrow -p_4$.  The diagrams on the left  will be denoted by $M_{9a}^{(1,2)}$ and those on the right  by $M_{9b}^{(1,2)}$ depending on whether the left vertex is $-{1 \over 2}\a^2\partial_\mu\b \partial^\mu\b$ (superscript (1)) or ${\omega^2 \over 2}\a^2\b^2$ (superscript (2)).}
 \label{L1abf}
\end{figure}
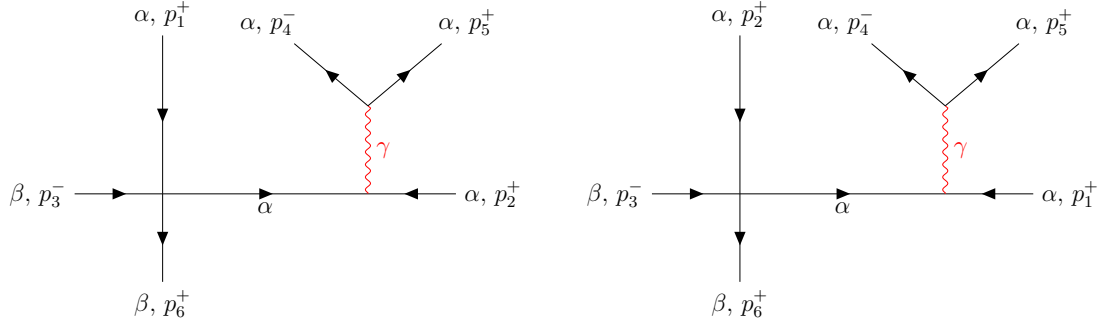
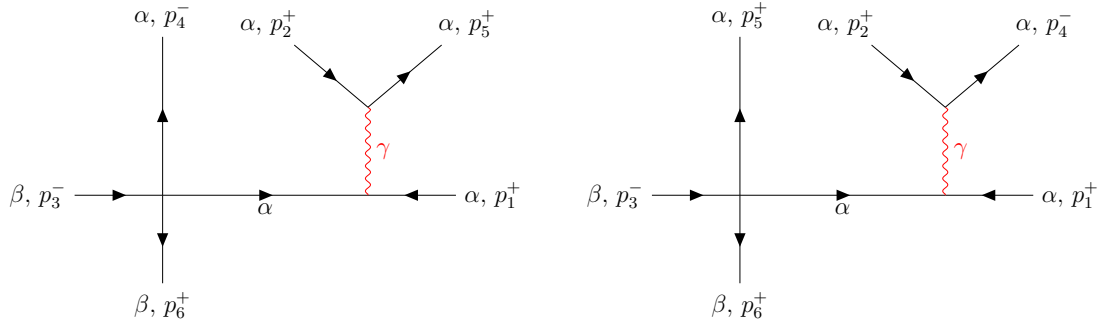
\begin{figure}

\resizebox{7.2cm}{!}{
\begin{tikzpicture}

\begin{feynman}
\vertex (a);
\vertex [below=of a](i1) {\(\b\), \( p_6^+\)};
\vertex [left=of a](i2) {\(\b\), \( p_3^-\)};
\vertex [above=2.7cm of a](f4){\(\a\), \( p_4^-\)};
\vertex [right=3.5cm of a](b);
\vertex [right=of b](f1){\(\a\), \( p_1^+\)}; 
\vertex [above=of b](c);
\vertex [above right=of c](f2){\(\a\), \( p_5^+\)};
\vertex [above left=of c](f3){\(\a\), \( p_2^+\)};

 \diagram* {
 [fermion] (a)--[fermion, edge label'=\(\a\)](b),
 (f1)--[fermion](b),
 (a)--[fermion](i1),
 (i2)--[fermion] (a),
 (a)--[fermion](f4),
 (b)-- [red,boson, edge label'=\(\gamma\)] (c)--[fermion](f2),
 (f3)--[fermion](c),
}; 

\hspace{280pt}

\vertex (a);
\vertex [below=of a](i1) {\(\b\), \( p_6^+\)};
\vertex [left=of a](i2) {\(\b\), \( p_3^-\)};
\vertex [above=2.7cm of a](f4){\(\a\), \( p_5^+\)};
\vertex [right=3.5cm of a](b);
\vertex [right=of b](f1){\(\a\), \( p_1^+\)}; 
\vertex [above=of b](c);
\vertex [above right=of c](f2){\(\a\), \( p_4^-\)};
\vertex [above left=of c](f3){\(\a\), \( p_2^+\)};

 \diagram* {
 [fermion] (a)--[fermion, edge label'=\(\a\)](b),
 (f1)--[fermion](b),
 (a)--[fermion](i1),
 (i2)--[fermion] (a),
 (a)--[fermion](f4),
 (b)-- [red,boson, edge label'=\(\gamma\)] (c)--[fermion](f2),
 (f3)--[fermion](c),
}; 
\end{feynman}

\end{tikzpicture}
}

\caption{Diagrams of type V with a non-zero contribution in the limit $\omega\rightarrow 0$ for the process  \eqref{a+b+b-->a-a+b+}.  One should add 4 additional diagrams of this type. Two like the left diagram, one with $p_1\leftrightarrow p_2$ and one with $p_1\leftrightarrow -p_5$ 
and two like the right diagram, one  with $p_1\leftrightarrow p_2$ and one with $p_1\leftrightarrow -p_4$. The diagrams on the left  will be denoted by $M_{9c}^{(1,2)}$ and those on the right  by $M_{9d}^{(1,2)}$ depending on whether the left vertex is $-{1 \over 2}\a^2\partial_\mu\b \partial^\mu\b$ (superscript (1)) or ${\omega^2 \over 2}\a^2\b^2$ (superscript (2)).}
 \label{L1cdf}
\end{figure}
We now turn to the last class of diagrams those depicted in fig. \ref{L1abf} and  \ref{L1cdf}.
The left vertex of the diagrams can be either $-{1 \over 2}\a^2\partial_\mu\b \partial^\mu\b$, in which case the corresponding contribution will be denoted by $M_9^{(1)}$, or ${\omega^2 \over 2}\a^2\b^2$, in which case the corresponding contribution will be denoted by $M_9^{(2)}$. In more detail, the contribution of the diagrams on the left of fig. \ref{L1abf} will be denoted by $M_{9a}^{(1,2)}$ and those on the right  by $M_{9b}^{(1,2)}$. Similarly, the diagrams on the  left of fig. \ref{L1cdf} will be denoted by $M_{9c}^{(1,2)}$ and those on the right  by $M_{9d}^{(1,2)}$. The corresponding contributions read
\begin{eqnarray}\label{L1a(1)}
M_{9a}^{(1)}=8 \omega^2 i \l^2 \Big(\tilde D(p_4+p_5)D(p_2-p_4-p_5)\,(p_4^0+p_5^0)^2\,p_3\cdot p_6+(p_2\leftrightarrow -p_4)+(p_2\leftrightarrow -p_5)\Big),\nonumber\\
\end{eqnarray}
where as always the notation $(p_2\leftrightarrow -p_5)$ means that one should add a term like the one written explicitly but with $p_2$ in the place of $-p_5$ and vice versa. \\
Moreover,
\begin{eqnarray}\label{L1b(1)}
M_{9b}^{(1)}=8 \omega^2 i \l^2 \Big(\tilde D(p_4+p_5)D(p_1-p_4-p_5)\,(p_4^0+p_5^0)^2\,p_3\cdot p_6+(p_1\leftrightarrow -p_4)+(p_1\leftrightarrow -p_5)\Big),\nonumber\\
\end{eqnarray}
\begin{eqnarray}\label{L1c(1)}
M_{9c}^{(1)}=8 \omega^2 i \l^2 \Big(\tilde D(p_2-p_5)D(p_1+p_2-p_5)\,(p_2^0-p_5^0)^2\,p_3\cdot p_6+(p_1\leftrightarrow p_2)+(p_1\leftrightarrow -p_5)\Big),\nonumber\\
\end{eqnarray}
\begin{eqnarray}\label{L1d(1)}
M_{9d}^{(1)}=8 \omega^2 i \l^2 \Big(\tilde D(p_2-p_4)D(p_1+p_2-p_4)\,(p_2^0-p_4^0)^2\,p_3\cdot p_6+(p_1\leftrightarrow p_2)+(p_1\leftrightarrow -p_4)\Big).\nonumber\\
\end{eqnarray}
Finally, one should sum to get
\be\label{L11}
M_{9}^{(1)}=M_{9a}^{(1)}+M_{9b}^{(1)}+M_{9c}^{(1)}+M_{9d}^{(1)}\, .
\ee
\begin{eqnarray}\label{L11}
M_{9}^{(1)}=16 i \l^2\frac{ p_2 p_4 p_5 p_6 p_1 \left(p_2^2-p_2 (p_5+p_6)+p_5 (p_5+p_6)\right)}{\omega^2 (p_2-p_5) (p_2-p_6) (p_5+p_6)}+\mathcal O(\omega^0).\nonumber\\
\end{eqnarray}
Notice that this result diverges in the limit $\omega\rightarrow 0$. This divergence cancels against the divergent part of $M_{10}^{(1)}$ which we compute in what follows.
Moreover, here we do not present the finite part of $M_{9}^{(1)}$ because it is quite lengthy. We will just mention that it also exactly cancels the finite part of $M_{10}^{(1)}$.

Similarly, when the left vertex is ${\omega^2 \over 2}\a^2\b^2$ one gets
\begin{eqnarray}\label{L1(2)}
&&M_{9c}^{(2)}=-8 \omega^4 i \l^2 \Big(\tilde D(p_2-p_5)D(p_1+p_2-p_5)\,(p_2^0-p_5^0)^2+(p_1\leftrightarrow p_2)+(p_1\leftrightarrow -p_5)\Big)\overset{\omega\rightarrow 0}{=}\nonumber\\
&&-8 i \l^2\frac{ p_2 p_5 p_1 \Big(p_2^2-p_2 (p_5+p_6)+p_5 (p_5+p_6)\Big)}{(p_2-p_5) (p_2-p_6) (p_5+p_6)}.
\end{eqnarray}
The other diagrams give zero in the massless limit, i.e.
\be\label{3zero}
M_{9a}^{(2)}=M_{9b}^{(2)}=M_{9d}^{(2)}=0.
\ee
So finally,
\be\label{L12}
M_{9}^{(2)}=M_{9c}^{(2)}.
\ee

\begin{figure}

\resizebox{7.2cm}{!}{
\begin{tikzpicture}

\begin{feynman}
\vertex (a);
\vertex [below=of a](i1) {\(\a\), \( p_5^+\)};
\vertex [left=of a](i2) {\(\b\), \( p_3^-\)};
\vertex [above=2.7cm of a](f4){\(\a\), \( p_4^-\)};
\vertex [right=3.5cm of a](b);
\vertex [right=of b](f1){\(\b\), \( p_6^+\)}; 
\vertex [above=of b](c);
\vertex [above right=of c](f2){\(\a\), \( p_2^+\)};
\vertex [above left=of c](f3){\(\a\), \( p_1^+\)};

 \diagram* {
 [fermion] (a)--[fermion, edge label'=\(\b\)](b),
 (b)--[fermion](f1),
 (a)--[fermion](i1),
 (i2)--[fermion] (a),
 (a)--[fermion](f4),
 (b)-- [red,boson, edge label'=\(\gamma\)] (c),
 (f2)--[fermion](c),
 (f3)--[fermion](c),
}; 

\hspace{280pt}

\vertex (a);
\vertex [below=of a](i1) {\(\a\), \( p_2^+\)};
\vertex [left=of a](i2) {\(\a\), \( p_4^-\)};
\vertex [above=2.7cm of a](f4){\(\a\), \( p_5^+\)};
\vertex [right=3.5cm of a](b);
\vertex [right=of b](f1){\(\a\), \( p_1^+\)}; 
\vertex [above=of b](c);
\vertex [above right=of c](f2){\(\b\), \( p_6^+\)};
\vertex [above left=of c](f3){\(\b\), \( p_3^-\)};

 \diagram* {
 [fermion] (a)--[fermion, edge label'=\(\a\)](b),
 (f1)--[fermion](b),
 (i1)--[fermion](a),
 (a)--[fermion] (i2),
 (a)--[fermion](f4),
 (b)-- [red,boson, edge label'=\(\gamma\)] (c)--[fermion](f2),
 (f3)--[fermion](c),
}; 
\end{feynman}

\end{tikzpicture}
}

\caption{Diagrams of type V  for the process  \eqref{a+b+b-->a-a+b+}. To the diagram on the left one should add 11 additional diagrams (see discussion right above \eqref{L21}). To the diagram on the right one should add three additional diagrams with the particles with momenta $p_1\leftrightarrow p_2$, $p_1\leftrightarrow -p_4$ and $p_1\leftrightarrow -p_5$ exchanged. These last 4 diagrams are 0 in the limit 
$\omega \rightarrow 0$.}
 \label{L23f}
\end{figure}
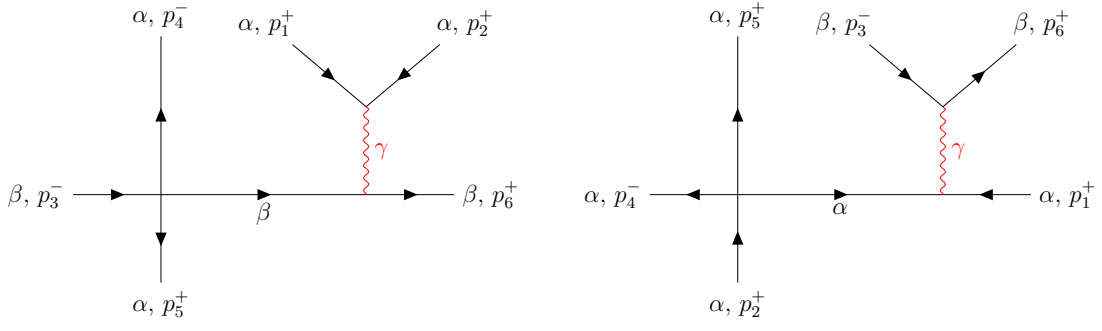

We now calculate the contribution of the diagrams on the left of fig. \ref{L23f}. As above, it will be denoted by $M_{10}^{(1)}$ when the left vertex is $-{1 \over 2}\a^2\partial_\mu\b \partial^\mu\b$ and by $M_{10}^{(2)}$ when the left vertex is ${\omega^2 \over 2}\a^2\b^2$. The full result will be obtained if to the contribution of the left diagram of fig. \ref{L23f} one adds eleven additional diagrams, one with 
the momenta $p_4\leftrightarrow -p_1$ exchanged, one with $p_4\leftrightarrow -p_2$, one with $p_5\leftrightarrow -p_1$, one with $p_5\leftrightarrow -p_2$ and one with $p_4\leftrightarrow -p_1 \land p_5\leftrightarrow -p_2$. Furthermore, to the six aforementioned terms one should add another six with $p_3\leftrightarrow -p_6$ exchanged. Adding up all terms one gets
\begin{eqnarray}\label{L21}
&&M_{10}^{(1)}=8 \omega^2 i \l^2 \Big(\tilde D(p_1+p_2)D(p_1+p_2-p_6)\,(p_1^0+p_2^0)^2\, p_3\cdot(p_3-p_4-p_5)+(11 \text{terms})\Big)
\nonumber\\
&&-16 i \l^2\frac{ p_2 p_4 p_5 p_6 p_1 \left(p_2^2-p_2 (p_5+p_6)+p_5 (p_5+p_6)\right)}{\omega^2 (p_2-p_5) (p_2-p_6) (p_5+p_6)}+\mathcal O(\omega^0),\nonumber\\
\end{eqnarray}
where again we have not written the finite term explicitly since it is lengthy and not very illuminating. The important point is that,  in the massless limit $\omega \rightarrow 0$, $M_{9}^{(1)}+M_{10}^{(1)}=0$.
Finally, for $M_{10}^{(2)}$ one obtains
\begin{eqnarray}\label{L1(2)}
&&M_{10}^{(2)}=-8 \omega^4 i \l^2 \Big(\tilde D(p_1+p_2)D(p_1+p_2-p_6)\,(p_1^0+p_2^0)^2+(11 \text{terms})\Big)\overset{\omega\rightarrow 0}{=}\nonumber\\
&&8 i \l^2\frac{ p_6 \left(p_6^2 \left(p_2^2+p_5^2\right)-2 p_6 (p_2-p_5) \left(p_2^2+p_5^2\right)+\left(p_2^2-p_2 p_5+p_5^2\right)^2\right)}{(p_2-p_5) (p_2-p_6) (p_5+p_6)}.
\end{eqnarray}
The ultimate contribution to the amplitude comes from the diagram on the right hand side of fig. \ref{L23f} accompanied by three additional diagrams with the particles with momenta $p_1\leftrightarrow p_2$, $p_1\leftrightarrow -p_4$ and $p_1\leftrightarrow -p_5$ exchanged.
By counting the number of powers of $\omega$ coming from the vertices and the propagators it is easily seen that each of these diagrams gives a zero result in the limit $\omega\rightarrow  0$.

Consequently, putting together all terms one gets a glorious zero
\be\label{equal-sets}
A_{\a^+\a^+\b^-\rightarrow\a^-\a^+\b^+} =\sum_{i=0}^{8} M_i+\sum_{i=9}^{10}\sum_{j=1}^{2}M_i^{(j)}= 0.
\ee Thus, we have shown that for generic kinematics the amplitude \eqref{a+b+b-->a-a+b+} is zero.
However, the calculation breaks down when the set of the momenta of the outgoing particles is equal to the set of the momenta of the incoming particles. This can be easily seen from \eqref{M1}, for example. It is obvious that the denominator of this expression explodes when $p_2=p_5$ or $p_2=p_6$. 
Needless to say that if the prescription of section \ref{DEF} is not followed the complete result for the amplitude is given by the sum of \eqref{M0} and \eqref{M1}, i.e. $M_0+M_1$ and  there is no constrain between 
the sets of the incoming and outgoing momenta, that is the amplitude \eqref{a+b+b-->a-a+b+} is non-zero for generic values of the incoming and outgoing momenta. If this could happen it would be in disagreement with the fact that the theory is integrable.

\subsection{A first  $3\rightarrow 3$ amplitude: factorisation of the S-matrix}
As mentioned in the previous section, the amplitude  \eqref{a+b+b-->a-a+b+} is zero unless $p_1=p_6,\,   p_2=p_5, \,p_3=p_4$ or $p_1=p_5, \, p_2=p_6, \,p_3=p_4$.
The uncertainty that arises in the expression \eqref{M1} for the aforementioned values of the momenta
is eliminated by rewriting the various propagators 
\be\label{princ}
D(p)={i \over p^2-m^2+i \epsilon}=i \,\mathcal P\big({1 \over p^2-m^2}\big)+\pi \,\delta(p^2-m^2).
\ee
The contribution of the principal values is canceled by that of the other diagrams considered in  section \ref{sec:6.1} since we have shown that the amplitude is zero \eqref{equal-sets} unless the sets of outgoing and incoming momenta are equal. As a result the total amplitude is written in the massless limit $\omega\rightarrow 0$ as

\begin{eqnarray}\label{M1fin}
&&A_{\a^+\a^+\b^-\rightarrow\a^-\a^+\b^+}=M_1\overset{\omega\rightarrow 0}{=}-2^2 \l^2 \pi \delta\big((p_2+p_3-p_4)^2\big)\,\,p_3\cdot p_2\,\,\,p_6\cdot p_2 
\nonumber \\
&&-2^2 \l^2  \pi \delta\big((p_1+p_3-p_4)^2\big)\,\,p_3\cdot p_1\, \,\, p_6\cdot p_1
 -2^2 \l^2 \pi \delta\big((p_2+p_3-p_5)^2\big)\,\,p_3\cdot p_3\,  \,\, p_6\cdot p_3
\nonumber \\
&&-2^2 \l^2 \pi \delta\big((p_3-p_4-p_5)^2\big)\,\,p_3\cdot p_4\,  \,\, p_6\cdot p_4 \nonumber \\
&&-2^2 \l^2  \pi \delta\big((p_1+p_2+p_3)^2\big)\,\,p_3\cdot (p_1+p_2)\,  \,\, p_6\cdot p_3
\nonumber \\
&&-2^2 \l^2  \pi \delta\big((p_1+p_3-p_5)^2\big)\,\,p_3\cdot p_1\,  \,\, p_6\cdot p_1=0.
\end{eqnarray}
Thus, we see that the amplitude vanishes because either the delta function is zero, as in the case of 
$\delta\big((p_1+p_2+p_3)^2\big)$, or because of the chosen kinematics which imply $p_6\cdot p_2=p_6\cdot p_1=0=p_3\cdot p_4=p_3\cdot p_3$.
Despite the fact that this amplitude is zero, it acquires the form of a sum of products of two $2\rightarrow 2$ amplitudes, as is dictated by integrability. It just happens that for the chosen kinematics one of the two $2\rightarrow 2$ amplitudes is zero nullifying the total amplitude.

\subsection{A second  $3\rightarrow 3$ amplitude: equality of the sets of initial and final momenta}
Next we focus on the following scattering amplitude
\be\label{a+a-b-->a+a-b-}
\a(p_1^+)\,  \a(p_2^-)\b(p_3^-)\,\rightarrow \a(p_4^+)\,\a(p_5^-)\, \b(p_6^-).
\ee
By considering all possible diagrams and performing a similar calculation with the one of  section \ref{sec:6.1} one can show that this amplitude vanishes 
\be\label{equal-sets-2}
A_{\a^+\a^-\b^-\rightarrow\a^+\a^-\b^-} = 0 \qq \text{unless} \qq
 \left.  \begin{array}{cc} p_1=p_4\\
           p_2=p_5\\
           p_3=p_6 \end{array}   \right \} \qq \text{or} \left. \qq \begin{array}{cc} p_1=p_4\\
           p_2=p_6\\
           p_3=p_5 \end{array}   \right \},
\ee
that is unless the sets of the incoming and outgoing momenta are equal.
Actually, the amplitude \eqref{a+a-b-->a+a-b-} is supported on $\delta$-functions imposing the equality of the sets of the momenta of the incoming and outgoing particles, as it should happen for an integrable theory.

In the next section we will argue that the amplitude \eqref{a+a-b-->a+a-b-} indeed factorises and can be written as  the sum of phases of two-body collisions.


\subsection{A second  $3\rightarrow 3$ amplitude: factorisation of the S-matrix}
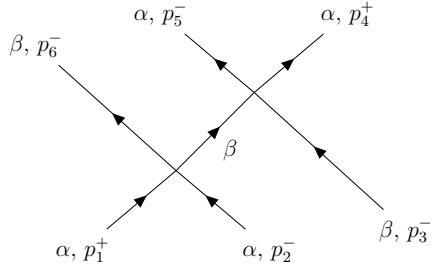
\begin{figure} 
\centering

\resizebox{6cm}{!}{
\begin{tikzpicture}


\begin{feynman}
\vertex (a);
\vertex [below right=of a](i1) {\(\alpha\), \( p_2^-\)};
\vertex [above left =2.7cm of a](f3) {\(\b\), \( p_6^-\)};
\vertex [below left=of a](f4) {\(\a\), \( p_1^+\)};
\vertex [above right=2cm of a](b);
\vertex [above left =of b](f2){\(\a\), \( p_5^-\)};
\vertex [above right=of b](f1){\(\a\), \( p_4^+\)};
\vertex [below right=3cm of b](i2){\(\b\), \( p_3^-\)};

 \diagram* {
(i1) -- [fermion] (a)-- [fermion, edge label'=\(\b\)] (b)-- [fermion] (f2),
(b)-- [fermion] (f1),
(i2) -- [fermion] (b),
(a) -- [fermion] (f3),
(f4) -- [fermion] (a),
}; 

\end{feynman}

\end{tikzpicture}
}
\caption{Diagram contributing to the amplitude \eqref{a+a-b-->a+a-b-} in the limit $\omega\rightarrow 0$. One should add 5 more diagrams of this type. One  with $p_1\leftrightarrow -p_4$,  one with $p_2\leftrightarrow -p_5$, one with $p_2\leftrightarrow -p_4$, one with $p_1\leftrightarrow -p_5$ and one with $p_1\leftrightarrow -p_4 \land p_2\leftrightarrow -p_5$ exchanged.}
 \label{M1pf}

\end{figure}
In this section we consider the amplitude  \eqref{a+a-b-->a+a-b-} and show that it factorises according to the requirements of integrability. The set of diagrams which give the final non-zero result is depicted in fig. \ref{M1pf}. Their contribution is
\begin{eqnarray}\label{M1p}
&&A_{\a^+\a^-\b^-\rightarrow\a^+\a^-\b^-}=-2^2 \l^2 D(p_1+p_2-p_6)\,\,p_3\cdot (p_1+p_2-p_6)\,\,\,\,  p_6\cdot (p_1+p_2-p_6) 
\nonumber \\
&&-2^2 \l^2 D(p_2-p_4-p_6)\,\,p_3\cdot (p_2-p_4-p_6)\, \,\,\, p_6\cdot (p_2-p_4-p_6)
\nonumber \\
&&-2^2 \l^2 D(p_1-p_5-p_6)\,\,p_3\cdot (p_1-p_5-p_6)\, \,\,\, p_6\cdot (p_1-p_5-p_6)
\nonumber \\
&&-2^2 \l^2 D(p_4+p_5+p_6)\,\,p_3\cdot (p_4+p_5+p_6)\, \,\,\, p_6\cdot (p_4+p_5+p_6)
\nonumber \\
&&-2^2 \l^2 D(p_1-p_4-p_6)\,\,p_3\cdot (p_1-p_4-p_6)\,\,\,\,  p_6\cdot (p_1-p_4-p_6)
 \\
&&-2^2 \l^2 D(p_2-p_5-p_6)\,\,p_3\cdot (p_2-p_5-p_6)\,  \,\,\,p_6\cdot (p_2-p_5-p_6).\nonumber
\end{eqnarray}
Now by applying \eqref{princ} and by remembering that the principal part has to cancel against the contribution of other diagrams since the amplitude under consideration is zero for generic kinematics one gets
\begin{eqnarray}\label{M1pfin}
&&A_{\a^+\a^-\b^-\rightarrow\a^+\a^-\b^-}\overset{\omega\rightarrow 0}{=}-2^2 \l^2 \pi\,  \delta\big((p_1+p_2-p_6)^2\big)\,\,p_3\cdot p_1\,  p_6\cdot p_1 
\nonumber \\
&&-2^2 \l^2  \pi \,\delta\big((p_2-p_4-p_6)^2\big)\,\,p_3\cdot p_4\, \,\, p_6\cdot p_4.
\end{eqnarray}
Before continuing let us note that the full contribution to the amplitude \eqref{a+a-b-->a+a-b-} comes solely from the diagrams of fig. \ref{M1pf}, as long as one imposes the equality of sets of the incoming and outgoing particles. This is so because all other diagrams give terms which are supported on $\delta$ functions -see \eqref{princ}- multiplied by positive powers of 
$\omega$ coming from the vertices, and as a result vanish in the limit $\omega \rightarrow 0$. The role of these diagrams is, of course  crucial for the vanishing of the amplitude when the momenta of the outgoing particles are not a permutation of the momenta of the incoming ones.

Returning to \eqref{M1pfin} , we see that the only possibility to get a non-zero result is when $p_1=p_4, \, p_2=p_6, \,p_3=p_5$. It is obvious now that the amplitude \eqref{a+a-b-->a+a-b-} is fully determined in terms of the 2-body scattering matrix and takes the factorised form
\begin{eqnarray}\label{M1pfinal}
&&A_{\a^+\a^-\b^-\rightarrow\a^+\a^-\b^-}=\\
&&A\Big(\a(p_1^+)\a(p_2^-)\rightarrow \b(p_1^+)\b(p_2^-)\Big) \pi\,  \delta\big((p_1+p_2-p_6)^2\big) A\Big(\b(p_1^+)\b(p_3^-)\rightarrow\a(p_1^+)\a(p_3^-)\Big)+
\nonumber \\
&&A\Big(\b(p_1^+)\a(p_2^-)\rightarrow\a(p_1^+)\b(p_2^-)\Big) \pi\,  \delta\big((p_2-p_4-p_6)^2\big) A\Big(\a(p_1^+)\b(p_3^-)\rightarrow\b(p_1^+)\a(p_3^-)\Big) \nonumber,
\end{eqnarray}
where the 4-point amplitudes $A\Big(\a(p_1^+)\a(p_2^-)\rightarrow\b(p_1^+)\b(p_2^-)\Big)$, $A\Big(\b(p_1^+)\b(p_3^-)\rightarrow\a(p_1^+)\a(p_3^-)\Big)$, $A\Big(\b(p_1^+)\a(p_2^-)\rightarrow\a(p_1^+)\b(p_2^-)\Big)$ and $A\Big(\a(p_1^+)\b(p_3^-)\rightarrow\b(p_1^+)\a(p_3^-)\Big)$ can be found in \eqref{aa>bb} and \eqref{ab>ab}.
Needless to say that the structure of \eqref{M1pfinal} is fully consistent with the restrictions of integrability.

\section{Conclusions}
In this work, we focused on the scattering matrix of integrable $\s$-models with massless excitations. 
In most of the cases, the corresponding scattering amplitudes suffer from IR issues. In particular, it often happens 
that for certain kinematic configurations some of the propagators go on-shell and, at the same time, one or both the vertices connected by these propagators vanish. As a result, the S-matrix is ill defined and its form depends on the precise way one regularises the aforementioned ambiguities. Even worse, in all schemes considered so far in the literature, integrability is broken and the massless S-matrix exhibits particle production and fails to factorise. 

In the present paper, we have proposed a definition of the S-matrix  of  integrable $\sigma$-models with massless excitations that is consistent with integrability. 
The key idea is to expand the Lagrangian of the theory around the non-trivial vacuum \eqref{nt-vac} depending on a single parameter $\omega$ which can eventually be sent to zero recovering in this way the massless theory.
One calculates all amplitudes by using  the Feynman rules derived from this Lagrangian and {\it only at the  end of the calculation} one sets $\omega=0$ in order to recover the massless theory. Despite the fact that our method works for integrable $\sigma$-models with one or more isometries, we believe that it can be applied even to the cases with no isometries at all, as long as one is able of finding and expanding around a non-trivial solution of the equations of motion that depends on a parameter which  can be  sent continuously to zero in order to obtain the trivial vacuum.
The S-matrix, so defined, does not support particle production, admits only equal sets of incoming and outgoing momenta and factorises.

At this point let us mention that the S-matrices of integrable $\sigma$-models around vacuum solutions of the type  \eqref{nt-vac} have been considered in the past for superstring models with a vanishing $\beta$-function \cite{Klose,Callan:2004uv}, as well as for  classical integrable $\sigma$-models with a non-vanishing $\beta$-function \cite{Georgiou:2022fow}.
There are two fundamental differences of these cases compared to the method employed in this work. 
Firstly, in all the cases considered in the past, one was imposing the Virasoro constraint in order to get rid of the non-physical degrees of freedom. Secondly, the parameter $\omega$ was never sent to zero. Consequently, the resulting S-matrices were explicitly breaking the Lorentz symmetry. In our construction the Lorentz symmetry, although broken in the intermediate steps of the calculation, is obeyed by the final result for the S-matrix (see, for example, \eqref{M1pfin}).

There are various directions for future research based on the present work. One direction is to apply the method to the scattering amplitudes at the loop level. It would be interesting to see if our prescription works at the quantum level, and if so, to clarify the precise way this happens. A second direction is to consider integrable $\sigma$-models with the $B$-field present. The  $B$-field will induce a mixing of the degrees of freedom even at the quadratic level and one should calculate amplitudes in the appropriate diagonal basis. A third direction would be to apply our prescription to more general integrable models. Good candidates for such an exploration is the 
$\l$-model \cite{Sfetsos:2013wia}, or even better the more general constructions of this type presented in \cite{Georgiou:2020wwg,Georgiou:2021pbd,Georgiou:2019nbz} which have the virtue of interpolating between different UV and IR fixed points.
Finally, and more ambitiously, one may use the present method to constrain and classify the space of massless integrable models by determining constraints on their various couplings order by order.

\subsection*{Acknowledgements}
We would like to thank Pantelis Panopoulos for useful discussions and encouragement during this work.






\end{document}